\documentclass[a4paper,notitlepage,twocolumn,footnoteinbib,longbibliography,superscriptaddress,10pt]{revtex4-1}

\usepackage{amsmath,amssymb,amsthm,mathtools,color,listings,graphicx}
\usepackage[colorlinks=true,linkcolor=blue,citecolor=blue]{hyperref}

\DeclareMathAlphabet{\mathbbb}{U}{bbold}{m}{n}

\begin{document}

\title{Equalized Hyperspin Machine}
\author{Marcello Calvanese Strinati}
\email{marcello.calvanesestrinati@gmail.com}
\affiliation{Centro Ricerche Enrico Fermi (CREF), Via Panisperna 89a, 00184 Rome, Italy}
\author{Claudio Conti}
\affiliation{Physics Department, Sapienza University of Rome, 00185 Rome, Italy}
\affiliation{Centro Ricerche Enrico Fermi (CREF), Via Panisperna 89a, 00184 Rome, Italy}
\date{\today}

\begin{abstract}
The reliable simulation of spin models is of critical importance to tackle complex optimization problems that are intractable on conventional computing machines. The recently introduced hyperspin machine, which is a network of linearly and nonlinearly coupled parametric oscillators, provides a versatile simulator of general classical vector spin models in arbitrary dimension, finding the minimum of the simulated spin Hamiltonian and implementing novel annealing algorithms. In the hyperspin machine,  oscillators evolve in time minimizing a cost function that must resemble the desired spin Hamiltonian in order for the system to reliably simulate the target spin model. This condition is met if the hyperspin amplitudes are equal in the steady state. Currently, no mechanism to enforce equal amplitudes exists. Here, we bridge this gap and introduce a method to simulate the hyperspin machine with equalized amplitudes in the steady state. We employ an additional network of oscillators (named equalizers) that connect to the hyperspin machine via an antisymmetric nonlinear coupling and equalize the hyperspin amplitudes. We demonstrate the performance of such an equalized hyperspin machine by large-scale numerical simulations up to $10000$ hyperspins. Compared to the hyperspin machine without equalization, we find that the equalized hyperspin machine (i) Reaches orders of magnitude lower spin energy, and (ii) Its performance is significantly less sensitive to the system parameters. The equalized hyperspin machine offers a competitive spin Hamiltonian minimizer and opens the possibility to combine amplitude equalization with complex annealing protocols to further boost the performance of spin machines.
\end{abstract}

\maketitle

\section{Introduction}
The study of the collective behaviour of spin systems is of paramount importance in modern science, both for fundamental and applied studies. In the last decade, a renewed interest in spin networks has been boosted by the possibility to map hard optimization problems onto specific spin Hamiltonians~\cite{10.3389/fphy.2014.00005}. Solving an optimization problem consists of finding the minimum state of a cost function defined by the interaction of a huge number of degrees of freedom. When mapped to a spin model, this translates into finding the spin configuration giving the minimum energy of the corresponding spin model. However, due to the exponentially large configuration space, the time required for the minimization scales exponentially with the system size. Such complexity currently pairs with the expected performance saturation of conventional computing architecture, causing complex combinatorial problems to become intractable on traditional hardware.

This circumstance has triggered a renewed interest in finding alternative computing paradigms to simulate spin models. To date, significant work has focused on the Ising model (binary spins) simulated by networks of driven dissipative nonlinear oscillators. This platform, named Ising machine, utilizes phase-dependent amplification to encode in the \emph{binary} phase of the oscillation the two states of an Ising spin~\cite{landauer1971,PhysRevA.88.063853}. Notable implementations include Bose-Einstein condensates~\cite{Byrnes_2011}, electrical (Kerr) oscillators~\cite{goto1959,PhysRevE.109.064308}, electro-optical systems~\cite{bohm2019}, digital Kerr simulators~\cite{hgoto2019cim,hgoto2021cim}, superconductive systems~\cite{goto2019kpoandopo,PhysRevA.93.050301}, and remarkably, degenerate optical parametric oscillators~\cite{PhysRevA.88.063853,marandi2014cim,nphoton.2016.68,s41534-017-0048-9,yamamoto2020isingmachine}. The last ones currently represent a valuable platform to simulate spin networks ensuring both reduced cost and room-temperature, ultrafast operation.

In the last years, continuous (vector) spin models have been also addressed. Most of the work has focused on the planar (XY) model. Compared to the Ising machines, the XY machine relies on phase-insensitive amplification and encodes in the \emph{continuous} phase of the amplified field the angle of the simulated XY spin. Proposed implementations of XY machines involve non-degenerate optical parametric oscillators~\cite{Takeda_2017}, laser networks~\cite{PhysRevResearch.2.043335,PhysRevResearch.2.033008,Gershenzon2020}, and polariton systems~\cite{kalininpolariton2017}.

The working mechanism of these systems is based on the fact that the dissipative dynamics of driven nonlinear oscillators, coupled via a linear coupling described by a matrix $\mathbf{J}$ defining the \emph{graph} of the network, behaves as a gradient descent: The oscillators undergo a dynamics where both their amplitudes and phases are self-regulated in time to minimize a Lyapunov (cost) function $L$, which accounts for individual oscillator energies as well as their mutual coupling~\cite{PhysRevLett.110.184102,Wang2017,arXiv:1903.07163,roychowdhury2022}. This kind of dynamics is referred to as \emph{unconstrained} gradient descent over $L$. Physically, the minimization of the Lyapunov function describes the tendency of the system to reach the state with minimal mode loss~\cite{PhysRevA.88.063853}.

In this framework, we recently introduced the hyperspin machine~\cite{strinati2022hyperspinmachine,PhysRevLett.132.017301}. Compared to the Ising and XY machines, which simulate either binary or planar spins, the hyperspin machine is a network of $D\times N$ parametric oscillators coupled both via linear and \emph{nonlinear} coupling. The nonlinear coupling, arising for example via common pump saturation~\cite{PhysRevLett.126.143901,Ben-Ami:23} where one pump field is depleted by $D$ oscillators at once, effectively organizes the $D\times N$ oscillators as $N$ vectors (hyperspins) where the $D$ parametric oscillators within a single hyperspin simulate the individual Cartesian components of each vector. The linear coupling encodes the graph of the network. Therefore, the hyperspin machine both unifies and generalizes the paradigm of using oscillator systems as spin simulators, providing a single system to simulate spins with tunable number of components $D$ (the Ising and XY spins are special cases of $D=1$ and $D=2$) and completely controllable coupling, ensured by the fact that spin vectors are defined via their individual Cartesian components.

Up to now, the hyperspin machine has been successfully tested both as minimizer of continuous spin ($D$-vector) models~\cite{RevModPhys.71.S358}, and as simulator of novel classical annealing protocols (\emph{dimensional annealing}) that effectively allow to interpolate during the time evolution between $D$-vector models at different dimension $D$. Inspired by quantum annealing~\cite{PhysRevE.58.5355}, dimensional annealing aims at approaching low-energy values of the simulated $D$-vector model by exploiting for an initial time transient the dynamics over a higher dimensional space $D'>D$. The idea of this protocol is that the dynamics of spins with $D'$ components ``softens'' the energy landscape of $D$-dimensional vectors by allowing the system to escape local minima more easily thanks to the higher dimension~\cite{strinati2022hyperspinmachine,PhysRevLett.132.017301}. As a consequence, the dynamics reaches an energy value of the simulated $D$-dimensional model that is closer to the actual global minimum compared to the straight simulation of the dynamics in $D$ dimensions. This protocol has been specifically tested by interpolating between $D=2$ and $D=1$, i.e., from XY to Ising model, to exponentially boost the performance of the system as an Ising machine. Later work based on a similar dimensionality reduction has been reported in Refs.~\cite{berloff2024visa,berloff2025covec,pierangeli2025polar}.

A critical issue when using parametric oscillators as spin simulators is that the presence of unconstrained amplitudes that converge to different values (\emph{amplitude heterogeneity}) causes the Lyapunov function $L$ to differ from the desired target one (Ising, XY, or in general $D$-vector model Hamiltonian). Generically speaking, this non-perfect mapping between $L$ and the target cost function due to amplitude heterogeneity represents a source of error: In Ising machines, this fact can severely reduce the probability that the system converges to the minimum energy state~\cite{PhysRevE.95.022118,PhysRevLett.122.040607,PhysRevLett.126.143901,roychowdhury2022}. In the hyperspin machine, amplitude heterogeneity causes the steady-state energy to be systematically larger than the global minimum of the simulated spin model, and it can be strongly dependent of the system parameters~\cite{strinati2022hyperspinmachine}.

To overcome this issue, previous work on Ising machines reported on the implementation of amplitude equalization by using a set of auxiliary variables (named ``error correction'' variables) whose goal is to enforce the oscillator amplitudes to be equal to a desired value in the steady state~\cite{PhysRevE.95.022118,PhysRevLett.122.040607}. Up to now, no such amplitude equalization mechanism exists for the hyperspin machine. The formulation of a method to suppress heterogeneity between hyperspin amplitudes represents a critical step in order to allow the hyperspin machine to work as a reliable simulator of vector spin models.

In this paper, we develop and validate on a large scale a method to suppress amplitude heterogeneity in the hyperspin machine, to realize what we name \emph{equalized hyperspin machine}. To introduce the reader to the concept of hyperspin machine, we first review in detail the construction of hyperspins from simple geometrical arguments, and then review the basic functioning of the hyperspin machine of Ref.~\cite{strinati2022hyperspinmachine}.

We then present the equalized hyperspin machine. Differently from the Ising machine, where one uses one error correction variable for each parametric oscillator to enforce equal amplitudes, heterogeneity in the hyperspin machine is corrected \emph{between} different hyperspins, but the parametric oscillators \emph{within} each hyperspin must remain unconstrained to preserve the vectorial nature of the simulated model. We show that this requirement is fulfilled by using a \emph{pair} of additional oscillators (named \emph{equalizers}, forming the \emph{equalization layer}) for each hyperspin. The equalizers couple to the parametric oscillators forming the hyperspin machine via an antisymmetric, nonlinear coupling.

We first demonstrate in great detail this method in the simplest case of two XY hyperspins, and then move to the general case of $N$, $D$-dimensional hyperspins. The main finding is that (i) The steady-state amplitude to which all hyperspins converge is automatically determined by the dynamics, and (ii) Its value is closely connected to the global minimum of the simulated $D$-vector Hamiltonian. 

The working mechanism of the equalized hyperspin machine is then validated by parallel numerical simulations on large-scale random graphs reaching up to $N=10000$, XY hyperspins, based on the formalism of nonlinear maps~\cite{strogatz2007nonlinear}, extending the numerical framework of Ref.~\cite{PhysRevLett.132.017301}. First, to scale up the system up to such large values of $N$, we show that it is necessary to perform a finite-size scaling of the coupling matrix $\mathbf{J}$ in order to ensure the existence of the steady state with equalized amplitudes. We discuss how the need of this scaling comes from the random nature of the considered graphs.

By comparing the minimum energy reached by the hyperspin machine with and without amplitude equalization, together with the hyperspin amplitude dynamics, we show that the suppression of amplitude heterogeneity makes the system reach significantly lower values of hyperspin energy compared to the unconstrained simulation.

Finally, we perform an extensive statistical analysis by comparing the performance of the hyperspin machine with and without amplitude equalization. We average on several random graphs the relative energy deviation, quantifying the difference between the energy found by the hyperspin machine and the actual minimum of the simulated model, and the degree of heterogeneity, quantifying amplitude heterogeneity between the hyperspins. For each graph, the relative energy deviation is computed as the difference between the minimum energy reached by the hyperspin machine and the value found by a python numerical minimizer.

This analysis shows that (i) The equalized hyperspin machine consistently finds orders of magnitude lower values of energy compared to the unconstrained hyperspin machine, (ii) The lower energy values come together with a lower heterogeneity degree, and (iii) The relative energy deviation does not significantly vary with $N$ even if the heterogeneity degree tends to increase with $N$. The last result indicates that the equalized hyperspin machine reliably converges very close to the global minimum even if amplitude equalization is not perfect.

A notable result that emerges from our analysis is that, for $N>400$, we find that the equalized hyperspin machine finds lower values of energy compared to python. This fact provides evidence that the equalized hyperspin machine offers competitive performance compared to existing method to minimize $D$-vector spin Hamiltonians.

This article is organized as follows. In Sec.~\ref{sec:parametricosicllatrosintrudoction01}, we recall the construction of hyperspins, first starting from the mapping between a single parametric oscillator and an Ising spin in Sec.~\ref{sec:parametricoscillators01}, and then moving to the mapping between $D$ nonlinearly coupled oscillators and one $D$-dimensional hyperspin in Sec.~\ref{sec:parametricoscillators02}. The unconstrained hyperspin machine and its working principle is recalled in Sec.~\ref{sec:parametricoscillators03}. Amplitude equalization is discussed from Sec.~\ref{sec:amplitudeequaizationinthehyperspinmachine01}, first for two XY hyperspins in Sec.~\ref{sec:examplecorrectionintwohyperspins01} and then for the equalized hyperspin machine for any $N$ and $D$ in Sec.~\ref{sec:examplecorrectionintwohyperspins02}. Finite-scaling of the coupling matrix to ensure the existence of the equalized steady state is discussed in Sec.~\ref{sec:examplecorrectionintwohyperspins03}. We discuss our numerical results in Sec.~\ref{sec:numericalretuls1}, and draw our conclusions in Sec.~\ref{sec:conclusions}.

\section{Hyperspins from degenerate parametric oscillators}
\label{sec:parametricosicllatrosintrudoction01}
Before reviewing the hyperspin machine, we open by recalling the construction of a single, $D$-dimensional hyperspin, explaining how $D$ nonlinearly coupled parametric oscillators can simulate a $D$-dimensional vector spin with arbitrary number $D$ of components.

\subsection{Parametric oscillator as Ising spin}
\label{sec:parametricoscillators01}
To proceed, it is instructive to first recall the reason why the field of a single, degenerate parametric oscillator (i.e., oscillating on a single-frequency mode) can be used to simulate a single, classical Ising spin. Generally speaking, a parametric oscillator (PO) consists of a resonator described by a real field $x(t)$, characterized by a proper frequency $\omega_0$ into which energy is injected by means of a driving field $p(t)=h\sin(\gamma t)$ with amplitude $h$ and frequency $\gamma$. When $\gamma$ is sufficiently close to $2\omega_0$, parametric resonance takes place: The system responds with an oscillation that occurs at $\gamma/2$, i.e., $x(t)\simeq|A|\cos(\gamma/2+\varphi)$. Because of the presence of the driving field, the phase of the oscillation $\varphi$ can take only two values: $0$ or $\pi$ with respect to the reference phase imposed by the drive. The value picked up by the oscillator depends on the initial conditions. This phenomenology, referred to as subharmonic response, or period-doubling instability, is reminiscent of the spontaneous $\mathbb{Z}_2$ (Ising) symmetry breaking, and allows to use the intrinsically binary nature of parametric oscillation to simulate a two-level system, such as a classical Ising spin (see Refs.~\cite{PhysRevA.100.023835,PhysRevLett.123.124301} and references therein). The two phase values identify the two states (``spin up'' and ``spin down'') of the Ising spin.

While this phenomenology is common to different nonlinear oscillators, most of the work discussed here is inspired by the optical parametric oscillator. The system consists of an optical cavity delimited by mirrors with a nonlinear medium inside. For concreteness, we focus on a second-order nonlinear medium in perfect phase-matching conditions~\cite{boyd2008nonlinear}. This medium is pumped by a strong laser field at frequency $\gamma=2\omega_0$, which amplifies degenerate signal and idler fields at frequency $\gamma/2=\omega_0$ and complex amplitude $A_1$ by spontaneous parametric down conversion, where the subscript ``$1$'' stresses that we now consider one oscillator field only. The pump field does not resonate with the cavity.

For a single optical field $A_1$, two main dynamical processes occur as the signal propagates inside the cavity: Parametric amplification with pump amplitude saturation and intrinsic mode loss. In proper dimensionless units, the resulting dynamics of $A_1$ is well captured by the following equation of motion~\cite{PhysRevA.43.6194,PhysRevA.88.063853,PhysRevA.109.063519}
\begin{equation}
\frac{dA_1}{dt}=\frac{h}{4}A^*_1-\frac{g}{2}A_1-\frac{\beta}{2}{|A_1|}^2A_1 \,\, ,
\label{eq:nonlineardynamicsparametricoscillator01}
\end{equation}
where the real quantities $h$, $g$, and $\beta$ denote respectively: Pump amplitude (parametric amplification), intrinsic loss coefficient (cavity losses), and nonlinear constant (pump saturation). The presence of $A^*_1$ in the right-hand side of Eq.~\eqref{eq:nonlineardynamicsparametricoscillator01} encapsulates phase-dependent amplification: The real part of $A_1$ is amplified while the imaginary part is suppressed in time. In the steady state, the field $A_1$ is therefore \emph{real} and the steady-state value is identified by the fixed points of Eq.~\eqref{eq:nonlineardynamicsparametricoscillator01}, found as customary by imposing $dA_1/dt=0$. At the steady state, one then has $|A_1|=\sqrt{(h/2-g)/\beta}\equiv S_1$. For a pump $h$ above the threshold value $h_{\rm th}=2g$, the parametric oscillator has two stable fixed points $A_1=\pm S_1$, symmetric with respect to the origin $A_1=0$.

\begin{figure*}[t]
\centering
\includegraphics[width=15.5cm]{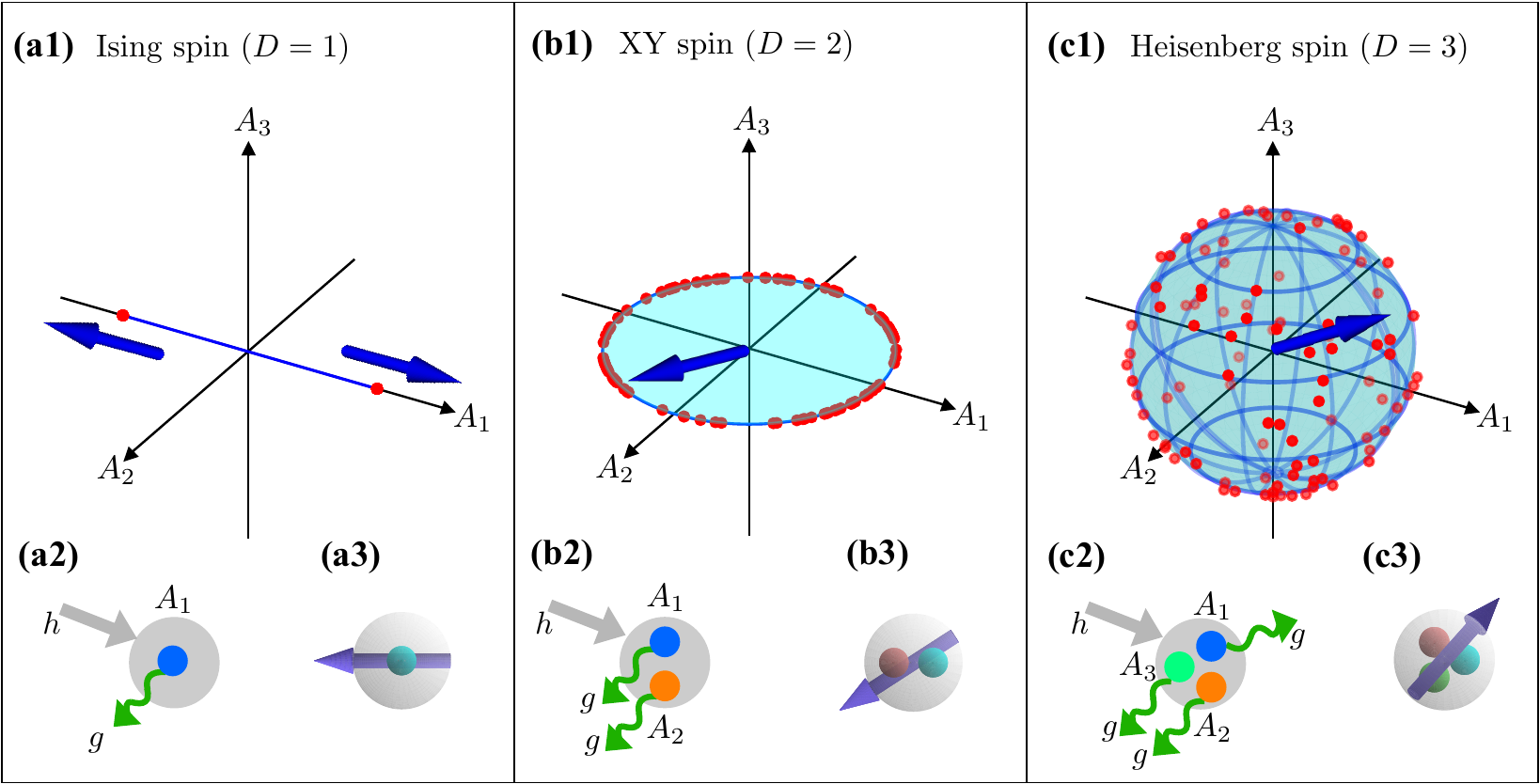}
\caption{Hyperspin fixed points from Eq.~\eqref{eq:nonlineardynamicsparametricoscillator02} with different $D$. \textbf{(a)} Ising spin $D=1$, \textbf{(b)} XY spin $D=2$, and \textbf{(c)} Heisenberg spin $D=3$. The fixed points are shown in the $(A_1,A_2,A_3)\subseteq\mathbb{R}^3$ space. For the Ising spin there is one oscillator only so $A_2=A_3=0$ in the figure. Similarly for the XY spin there are two oscillators, so $A_3=0$. The red dots are the fixed points of the dynamics obtained by integrating Eq.~\eqref{eq:nonlineardynamicsparametricoscillator02} several times for different initial conditions. The fixed points are found on the boundary of a $D$-dimensional surface (\emph{hypersphere}), which in particular is a segment, a circle, and a sphere for $D=1,2,3$, respectively. The oscillator amplitudes can then be used to define a $D$-dimensional vector (blue arrow), which simulates a $D$-dimensional spin (\emph{hyperspin}). Bottom panels depict the nonlinearly coupled oscillators (left panels) and the corresponding representation as hyperspin vectors $\vec{\sigma}_1=\vec{S}_1/S_1$ (right panels), for the same $D$ as in the corresponding top panel. Colored dots depict the individual oscillator amplitudes, subject to intrinsic loss (quantified by $g$, green wavy arrows), and the outer gray circle with gray arrow depicts the pump ($h$) that is saturated by all the oscillators at once. The geometrical interpretation is here shown for $D=1,2,3$ for illustration purposes, but it is valid for any $D$ (see Ref.~\cite{strinati2022hyperspinmachine} for an extended discussion).}
\label{fig:sketchfixedpoints01}
\end{figure*}

Another way to see that Eq.~\eqref{eq:nonlineardynamicsparametricoscillator01} admits two solutions is by noting that the equation under the transformation $A_1\rightarrow e^{i\varphi}A_1$ is invariant only if $\varphi=0$ or $\varphi=\pi$ because of the fact that the right-hand side of the equation depends on both $A_1$ and $A^*_1$. In our specific case where $A_1$ converges to a real value, then $\sigma_1={\rm sign}(A_1)$ simulates an Ising spin (see Fig.~\ref{fig:sketchfixedpoints01}\textbf{a} for a pictorial representation).

\subsection{Extending the dimensionality of the spin}
\label{sec:parametricoscillators02}
The two fixed points of the parametric oscillator can be seen as the boundary of the one-dimensional space $I\subseteq\mathbb{R}$ (i.e., a segment) over which the amplitude $A_1$ in Eq.~\eqref{eq:nonlineardynamicsparametricoscillator01} is defined. As said in Sec.~\ref{sec:parametricoscillators01}, the presence of \emph{two} fixed points is what allows to use the phase of a parametric oscillator to simulate an Ising spin $\sigma_1$. The concept of ``hyperspin'' introduced in Ref.~\cite{strinati2022hyperspinmachine} aims at extending the applicability of parametric oscillators, generalizing the dynamics to yield a steady state that allows the simulation of a generic vector spin $\vec{\sigma}_1$ with arbitrary number $D$ of components.

Crucially, we demand that such a vector spin is \emph{defined} by its individual Cartesian components. To achieve this, we see that a single hyperspin needs to be defined by $D$ parametric oscillators, with amplitudes $A_j$ forming a vector $\vec{S}_1=(A_1,\ldots,A_D)$. The Ising spin is the special case of $D=1$. Following the discussion in Sec.~\ref{sec:parametricoscillators01}, we now focus on real amplitudes, since the imaginary part of the amplitudes are suppressed during the time evolution.

The key requirement that lies at the core of the hyperspin design is that the dynamical system of oscillators has fixed points that are found on the boundary of a $D$-dimensional space $I\subseteq\mathbb{R}^D$. Basically, we construct the single hyperspin dynamics such that the steady state is defined on a continuous manifold of dimension $D-1$ and not just by two points. One possible way to achieve this is by modifying the nonlinear term in Eq.~\eqref{eq:nonlineardynamicsparametricoscillator01} as
\begin{equation}
\frac{dA_j}{dt}=\frac{h}{4}A_j-\frac{g}{2}A_j-\frac{\beta}{2}\left(\sum_{r=1}^{D}A^2_r\right)A_j \,\, ,
\label{eq:nonlineardynamicsparametricoscillator02}
\end{equation}
or equivalently in the vector notation~\cite{PhysRevLett.126.143901}
\begin{equation}
\frac{d\vec{S}_1}{dt}=\left(\frac{h}{4}-\frac{g}{2}-\frac{\beta}{2}{\left|\!\left|\vec{S}_1\right|\!\right|}^2\right)\vec{S}_1 \,\, ,
\label{eq:nonlineardynamicsparametricoscillator02bis01}
\end{equation}
where $|\!|\vec{S}_1|\!|^2\equiv\vec{S}_1\cdot\vec{S}_1\equiv S^2_1=\sum_{r=1}^{D}A^2_r$ is the squared norm of $\vec{S}_1$ and `` $\cdot$ '' denotes the standard dot (Euclidian scalar) product. Indeed, one can see that the fixed-point condition $dA_j/dt=0$ with $A_j\neq0$ for all $j$ imposes that
\begin{equation}
\frac{h}{2}-g-\beta\sum_{r=1}^{D}A^2_r=0 \,\, ,
\label{eq:nonlineardynamicsparametricoscillator03}
\end{equation}
which means that the parametric oscillator amplitudes at the steady state satisfy the constraint
\begin{equation}
\sum_{r=1}^{D}A^2_r=\frac{1}{\beta}\left(\frac{h}{2}-g\right)\equiv S^2_1 \,\, .
\label{eq:nonlineardynamicsparametricoscillator04}
\end{equation}
The fixed points are then defined on a $D$-dimensional hypersphere of radius $S_1$. Compared to Eq.~\eqref{eq:nonlineardynamicsparametricoscillator01}, the nonlinearity in Eq.~\eqref{eq:nonlineardynamicsparametricoscillator02} is \emph{nonlocal} (or \emph{global}) and effectively induces a nonlinear coupling between the oscillators, i.e., $A_j$ is coupled to all other oscillators by cubic coupling terms of the form $A^2_rA_j$. This nonlinear coupling between parametric oscillators inducing the hyperspin structure can be seen arising in two possible ways:
\begin{itemize}
\item As a nonlocal pump saturation, i.e., $h$ is saturated by each oscillator $A_j$ as $h-2\beta\sum_{r=1}^{D}A^2_r$;
\item As a nonlocal intrinsic loss, i.e., $g$ for each oscillator $A_j$ is modified as $g+\beta\sum_{r=1}^{D}A^2_r$.
\end{itemize}
This suggests two possible ways to realize hyperspins by the effective nonlinear coupling. In Ref.~\cite{strinati2022hyperspinmachine} and Fig.~\ref{fig:sketchfixedpoints01} hyperspins are formulated considering a nonlocal pump saturation~\cite{PhysRevLett.126.143901,Ben-Ami:23}, while in Ref.~\cite{PhysRevLett.132.017301} hyperspins are induced via a nonlocal loss.

We remark that the dynamics in Eq.~\eqref{eq:nonlineardynamicsparametricoscillator02} does not fix the amplitudes $A_j$ individually, but only their squared sum in Eq.~\eqref{eq:nonlineardynamicsparametricoscillator04}. As a consequence, the point on the hypersphere to which the oscillator amplitudes converge depends on the initial conditions. The components of the hyperspin vector are the individual oscillator amplitudes, which determine the hyperspin direction and orientation.

Equation~\eqref{eq:nonlineardynamicsparametricoscillator02bis01} is invariant under the transformation $\vec{S}_1\rightarrow\mathbf{R}\,\vec{S}_1$, where $\mathbf{R}$ a generic rotation matrix ($\mathbf{R}$ is an isometry that leaves $\vec{S}_1\cdot\vec{S}_1$ invariant since $\mathbf{R}^T\mathbf{R}=\mathbbb{1}$). Correspondingly, we define the $D$-dimensional vector spin as $\vec{\sigma}_1=\vec{S}_1/S_1$.

Figure~\ref{fig:sketchfixedpoints01}\textbf{b} pictorially explains the mapping between the fixed points and the $D$-dimensional spin vector for $D=2$ (vector defined on the $xy$-plane, also known as planar or XY spin). Figure~\ref{fig:sketchfixedpoints01}\textbf{c} refers to the case $D=3$ (vector in the $xyz$-space, also known as Heisenberg spin). Red dots are the fixed points obtained by integrating Eq.~\eqref{eq:nonlineardynamicsparametricoscillator02} for different initial conditions. The fixed points are found on a circumference and on a sphere of radius $S_1$ [Eq.~\eqref{eq:nonlineardynamicsparametricoscillator04}] for $D=2$ and $D=3$, respectively. Accordingly, from the oscillator amplitudes, the XY spin in Fig.~\ref{fig:sketchfixedpoints01}\textbf{b} is defined as $\vec{\sigma}_1=(A_1,A_2)/\sqrt{A^2_1+A^2_2}$, while the Heisenberg spin in Fig.~\ref{fig:sketchfixedpoints01}\textbf{c} is defined as $\vec{\sigma}_1=(A_1,A_2,A_3)/\sqrt{A^2_1+A^2_2+A^2_3}$. Figure~\ref{fig:sketchfixedpoints01} gives examples for $D=1,2,3$ to ease their geometrical interpretation. We remark that the hyperspin construction is general and valid at any $D$~\cite{strinati2022hyperspinmachine}.

\begin{figure}[t]
\centering
\includegraphics[width=7cm]{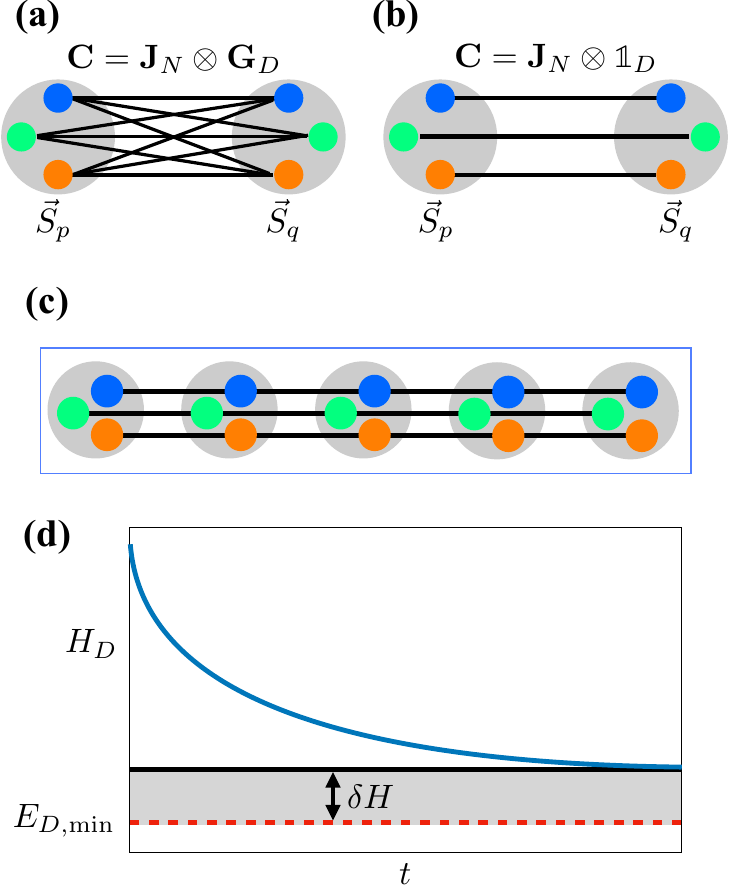}
\caption{\textbf{(a)} Scheme of the linear coupling matrix $\mathbf{C}$ between any two hyperspins $\vec{S}_q$ and $\vec{S}_p$ with generic $\mathbf{G}_D$ [Eqs.~\eqref{eq:nonlineardynamicsparametricoscillator08} and~\eqref{eq:nonlineardynamicsparametricoscillator09}], with $D=3$ for illustration purposes. Hyperspins are represented as in Fig.~\ref{fig:sketchfixedpoints02} and the linear coupling connections are depicted by the black lines between the colored dots. \textbf{(b)} As in panel \textbf{(a)} but with $\mathbf{G}_D=\mathbbb{1}_D$, which implements the standard Euclidian metric. Graphically, only the connections between dots of the same color (which are the homologous vector components between the two hyperspins) are nonzero. \textbf{(c)} Representation of the hyperspin machine used throughout this paper, obtained by connecting hyperspins with arbitrary $\mathbf{J}_N$. \textbf{(d)} Scheme of the working principle of the hyperspin machine with unconstrained hyperspin amplitudes. The $D$-vector model energy $H_D$ [blue line, from Eq.~\eqref{eq:dynamicsamplitueheterogeneity07}] computed from the hyperspin amplitudes in time decreases and reaches a steady-state value that can be significantly larger than the actual global minimum $E_{D,{\rm min}}$ [red dashed line, from Eq.~\eqref{eq:amplitudehomogeneityhyperpsin18bis7bis1}].}
\label{fig:sketchfixedpoints02}
\end{figure}

\section{The hyperspin machine}
\label{sec:parametricoscillators03}
\subsection{Nonlinear and linear coupling}
In this section, we review the construction of the hyperspin machine by linearly and nonlinearly coupled parametric oscillators proposed in Ref.~\cite{strinati2022hyperspinmachine}. A system of $N$ coupled hyperspins consists of a network of $N\times D$ parametric oscillators, which are organized as $N$, $D$-dimensional hyperspins by the effective nonlinear coupling as in Eq.~\eqref{eq:nonlineardynamicsparametricoscillator02}. Considering only the nonlinear coupling, the dynamics of $A_j$ is described by the equations of motion ($j=1,\ldots,DN$)
\begin{equation}
\frac{dA_j}{dt}=\frac{h}{4}A_j-\frac{g}{2}A_j-\frac{1}{2}A_j\left(\sum_{r=1}^{DN}W_{jr}\,A^2_r\right) \,\, ,
\label{eq:nonlineardynamicsparametricoscillator05}
\end{equation}
where the $DN\times DN$ symmetric matrix $\mathbf{W}$ expresses the nonlinear coupling between parametric oscillators inducing the hyperspins. By construction, $\mathbf{W}$ is the block diagonal matrix written as
\begin{equation}
\mathbf{W}=\beta\,\mathbbb{1}_N\otimes\mathcal{I}_D \,\, ,
\label{eq:nonlineardynamicsparametricoscillator06}
\end{equation}
where $\mathbbb{1}_N$ is the $N\times N$ identity matrix, $\mathcal{I}_D$ is the $D\times D$ matrix of ones, and ``$\otimes$'' denotes the Kr\"{o}necker product. We define the hyperspin vector as $\vec{S}_q=(A_{1+(q-1)D},\ldots,A_{qD})$, with $q=1,\ldots,N$. Accordingly, Eq.~\eqref{eq:nonlineardynamicsparametricoscillator06} yields $N$ decoupled equations of the form of Eq.~\eqref{eq:nonlineardynamicsparametricoscillator02bis01}, i.e.
\begin{equation}
\frac{d\vec{S}_q}{dt}=\left(\frac{h}{4}-\frac{g}{2}-\frac{\beta}{2}{\left|\!\left|\vec{S}_q\right|\!\right|}^2\right)\vec{S}_q \,\, ,
\label{eq:nonlineardynamicsparametricoscillator07}
\end{equation}
where $|\!|\vec{S}_q|\!|^2\equiv\vec{S}_q\cdot\vec{S}_q\equiv S^2_q=\sum_{r=1}^{D}A^2_{r+(q-1)D}$ is the squared norm of the $q$-th hyperspin. 

Then, the individual parametric oscillators are also connected by a dissipative linear coupling, expressed by symmetric matrix $\mathbf{C}$ with dimension $DN\times DN$. This induces a linear coupling between different components of distinct spin vectors. Specifically, by writing
\begin{equation}
\mathbf{C}=\mathbf{J}_N\otimes\mathbf{G}_D \,\, ,
\label{eq:nonlineardynamicsparametricoscillator07bis01}
\end{equation}
the linear coupling matrix accounts for two terms:
\begin{itemize}
\item The $N\times N$ \emph{graph} matrix $\mathbf{J}_N$ with elements $J_{pq}$ and zero diagonal entries, expressing the coupling weight between any two hyperspins $\vec{S}_q$ and $\vec{S}_p$;
\item The $D\times D$ \emph{metric} matrix $\mathbf{G}_D$, expressing for any two hyperspins $\vec{S}_q$ and $\vec{S}_p$ how the coupling weight $J_{pq}$ is distributed between the individual oscillators forming $\vec{S}_q$ and $\vec{S}_p$.
\end{itemize}
The fact that $\mathbf{J}_N$ has zero diagonal entries means that parametric oscillators within the same hyperspin are not coupled by linear coupling (see Fig.~\ref{fig:sketchfixedpoints02}\textbf{a} for a sketch of the coupling with $D=3$).

With linear coupling, Eq.~\eqref{eq:nonlineardynamicsparametricoscillator05} reads
\begin{equation}
\frac{dA_j}{dt}=\frac{h}{4}A_j-\frac{g}{2}A_j-\frac{1}{2}A_j\left(\sum_{r=1}^{DN}W_{jr}\,A^2_r\right)+\frac{1}{2}\sum_{k=1}^{DN}C_{jk}A_k \, .
\label{eq:nonlineardynamicsparametricoscillator08}
\end{equation}
The hierarchical nature of the linear coupling becomes evident by explicitly expanding the Kr\"{o}necker product as follows~\cite{strinati2022hyperspinmachine}. If  $A_j\equiv S^{(\mu)}_q$ defines the $\mu$-th component to the $q$-th hyperspin $\vec{S}_q$, i.e., $q=1+\lfloor(j-1)/D\rfloor$ and $\mu=1+(j-1){\rm mod}(D)$, where $\lfloor\cdot\rfloor$ and ${\rm mod}$ are the floor and modulo functions, respectively, the coupling term can be also written as
\begin{equation}
\sum_{k=1}^{DN}C_{jk}A_k=\sum_{p=1}^{N}J_{pq}\sum_{\nu=1}^{D}G_{\mu\nu}\,A_{\nu+(p-1)D} \,\, .
\label{eq:nonlineardynamicsparametricoscillator09}
\end{equation}

\subsection{Lyapunov function}
Hereafter, we focus on Euclidian metric, obtained by setting $\mathbf{G}_D=\mathbbb{1}_D$, depicted in Fig.~\ref{fig:sketchfixedpoints02}\textbf{b}. In this case, Eq.~\eqref{eq:nonlineardynamicsparametricoscillator08} takes the vector form
\begin{equation}
\frac{d\vec{S}_q}{dt}=\left(\frac{h}{4}-\frac{g}{2}-\frac{\beta}{2}{\left|\!\left|\vec{S}_q\right|\!\right|}^2\right)\vec{S}_q+\frac{1}{2}\sum_{p=1}^{N}J_{pq}\,\vec{S}_p \,\, .
\label{eq:nonlineardynamicsparametricoscillator10}
\end{equation}
The dynamics in Eqs.~\eqref{eq:nonlineardynamicsparametricoscillator08} and~\eqref{eq:nonlineardynamicsparametricoscillator10} defines the hyperspin machine for any graph matrix $\mathbf{J}_N$. As in Eq.~\eqref{eq:nonlineardynamicsparametricoscillator02bis01}, Eq.~\eqref{eq:nonlineardynamicsparametricoscillator10} is invariant under the transformation $\vec{S}_q\rightarrow\mathbf{R}\,\vec{S}_q$, where $\mathbf{R}$ the same generic rotation matrix for all $q$. Figure~\ref{fig:sketchfixedpoints02}\textbf{c} depicts the hyperspin machine in the specific case of Euclidian metric, implemented by Eq.~\eqref{eq:nonlineardynamicsparametricoscillator10}.

One can show that the dynamics in Eq.~\eqref{eq:nonlineardynamicsparametricoscillator08} with $\mathbf{C}=\mathbf{J}_N\otimes\mathbbb{1}_D$ is obtained from the following Lyapunov function $L=L(A_1,\ldots,A_{DN})$ that accounts for both individual hyperspin terms and mutual coupling between all hyperspins~\cite{strinati2022hyperspinmachine}
\begin{eqnarray}
L&=&-\frac{1}{2}\sum_{q=1}^{N}\left[\left(\frac{h}{4}-\frac{g}{2}\right)S^2_q-\frac{\beta}{4}S^4_q\right]-\frac{1}{4}\sum_{p,q=1}^{N}J_{pq}\,\vec{S}_p\cdot\vec{S}_q\nonumber\\
&=&L_0+L_{\rm coupl} \,\, ,
\label{eq:nonlineardynamicsparametricoscillator11}
\end{eqnarray}
as
\begin{equation}
\frac{dA_j}{dt}=-\frac{\partial L}{\partial A_j} \,\, .
\end{equation}
This means that the dynamics is such that
\begin{equation}
\frac{dL}{dt}=\sum_{j=1}^{DN}\frac{\partial L}{\partial A_j}\,\frac{dA_j}{dt}=-\sum_{j=1}^{DN}{\left(\frac{dA_j}{dt}\right)}^2\leq0 \,\, .
\label{eq:nonlineardynamicsparametricoscillator11bis01}
\end{equation}
In Eq.~\eqref{eq:nonlineardynamicsparametricoscillator11}, $L_0$ includes all the first summation over $q$ (individual hyperspin terms) and $L_{\rm coupl}$ is the second summation over $p$ and $q$ (hyperspin coupling).

The inequality $dL/dt\leq 0$ in Eq.~\eqref{eq:nonlineardynamicsparametricoscillator11bis01} encapsulates the working principle of the hyperspin machine, which generalizes that of the Ising machine~\cite{roychowdhury2022}, and it can be summarized as follows: The nonlinear dynamics of linearly ($\mathbf{C}$) and nonlinearly ($\mathbf{W}$) coupled parametric oscillators in Eq.~\eqref{eq:nonlineardynamicsparametricoscillator08} behaves as a gradient descent tending to minimize the Lyapunov function in Eq.~\eqref{eq:nonlineardynamicsparametricoscillator11} that, \emph{when the hyperspin amplitudes are equal}, i.e., $S_q\equiv S$, for all $q$, becomes proportional to the $D$-vector model energy $L=L_0+S^2H_D/4$. The $D$-vector model energy is defined as~\cite{RevModPhys.71.S358}
\begin{equation}
H_D(\{\vec{\sigma}_q\})=-\sum_{p,q=1}^{N}J_{pq}\,\vec{\sigma}_p\cdot\vec{\sigma}_q \,\, .
\label{eq:dynamicsamplitueheterogeneity07}
\end{equation}
Notice that in Eq.~\eqref{eq:nonlineardynamicsparametricoscillator08} there is no process that constrains the hyperspin amplitudes to be equal. In other words, the dynamics in Eq.~\eqref{eq:nonlineardynamicsparametricoscillator08} performs unconstrained optimization on the Lyapunov function $L$. When amplitude equalization is not perfect, i.e., $S_q=(1+\Delta_q)S$ for some heterogeneity $\Delta_q$, then accordingly one has
\begin{equation}
L=L_0+\frac{S^2}{4}\,H_D+\delta L \,\, ,
\label{eq:dynamicsamplitueheterogeneity07bis0}
\end{equation}
where $\delta L$ accounts for all terms proportional to $\Delta_q$. As in the Ising machine~\cite{PhysRevE.95.022118,PhysRevLett.122.040607,Vadlamani26639}, the presence of non-equalized amplitudes reflects into a Lyapunov function that is not properly mapped onto the target cost function, which for the hyperspin machine is Eq.~\eqref{eq:dynamicsamplitueheterogeneity07}.

This fact is depicted in Fig.~\ref{fig:sketchfixedpoints02}\textbf{d}. The solid blue line sketches the hyperspin energy in time from Eq.~\eqref{eq:dynamicsamplitueheterogeneity07}, using the spin variables $\vec{\sigma_q}=\vec{S}_q/S_q$ with $\vec{S}_q$ defined from the oscillator amplitudes in Eq.~\eqref{eq:nonlineardynamicsparametricoscillator08}. As found in Refs.~\cite{strinati2022hyperspinmachine,PhysRevLett.132.017301}, the hyperspin energy decreases until it reaches a steady-state value (black solid horizontal line in Fig.~\ref{fig:sketchfixedpoints02}\textbf{d}) that depends on the system parameters, crucially the pump amplitude $h$. For hard graph connectivities $\mathbf{J}_N$, the steady-state hyperspin energy can be higher than the global minimum $E_{D,{\rm min}}$ of Eq.~\eqref{eq:dynamicsamplitueheterogeneity07}
\begin{equation}
E_{D,{\rm min}}\coloneqq\min_{\{\vec{\sigma}_q\}}H_D(\{\vec{\sigma}_q\})<0 \,\, ,
\label{eq:amplitudehomogeneityhyperpsin18bis7bis1}
\end{equation}
depicted in Fig.~\ref{fig:sketchfixedpoints02}\textbf{d} as the red dashed horizontal line. As pointed out in Ref.~\cite{strinati2022hyperspinmachine}, the presence of a nonzero energy difference $\delta H>0$ between the long-time hyperspin energy and the global minimum of $H_D$ (shaded area in Fig.~\ref{fig:sketchfixedpoints02}\textbf{d}) is related to the presence of non-equal hyperspin amplitudes, thereby giving a $\delta L\neq0$ in the Lyapunov function [Eq.~\eqref{eq:dynamicsamplitueheterogeneity07bis0}]. The energy difference $\delta H$ decreases as amplitudes become equalized~\cite{strinati2022hyperspinmachine}. When amplitudes are exactly equal, $\delta L$ and $\delta H$ vanish.

An open challenge is the realization of a mechanism to equalize amplitudes in the hyperspin machine to reach $E_{D,{\rm min}}$ independently of the system parameters. The rest of this paper is devoted to fill this gap, by proposing and validating a method for amplitude equalization.


%

\section{Amplitude equalization in the hyperspin machine}
\label{sec:amplitudeequaizationinthehyperspinmachine01}
In this section, we discuss our method for amplitude equalization. First we introduce the general principle, and then we show how to implement the scheme via nonlinear maps. We introduce an additional layer of $M=2(N-1)$ oscillators (hereafter, the \emph{equalizers}) connected to the POs forming the hyperspin machine in Fig.~\ref{fig:sketchfixedpoints02}\textbf{c}. The role of the equalizers is to ``absorb'' amplitude heterogeneity in the hyperspin machine outlined in Sec.~\ref{sec:parametricoscillators03}, such that the amplitudes are equal at the steady state.

\subsection{Equalization in $N=2$ hyperspins}
\label{sec:examplecorrectionintwohyperspins01}
We first focus on the simplest case of $N=2$ hyperspins, coupled via a symmetric linear coupling $J_{12}=J_{21}=c$. For two hyperspins, amplitude heterogeneity can be induced, e.g., by different pump amplitudes $h_1$ and $h_2$.

\begin{figure}[t]
\centering
\includegraphics[width=8.4cm]{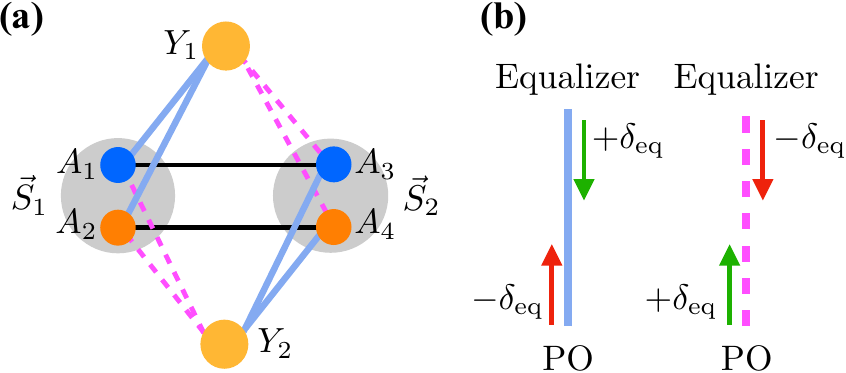}
\caption{\textbf{(a)} Connectivity of two hyperspins formed by parametric oscillators $A_{j}$ (colored dots in the gray circles) to two auxiliary oscillators $Y_{k}$ named \emph{equalizers} (yellow dots) performing amplitude equalization. Here $N=2$ and $D=2$ (XY hyperspins) are chosen for illustration purposes. The oscillator amplitudes $A_1$, $A_2$, $A_3$, and $A_4$ define two XY vectors $\vec{S}_1=(A_1,A_2)$ and $\vec{S}_2=(A_3,A_4)$, coupled by a linear coupling implementing the Euclidian scalar (dot) product $\vec{S}_1\cdot\vec{S}_2$ (black lines). The PO amplitudes $A_j$ are coupled by a directional, antisymmetric nonlinear coupling (see Table~\ref{table:couplingandtypeofcouplinghyperspinmachine01}) with coupling constant $\delta_{\rm eq}>0$ to the two equalizers $Y_1$ and $Y_2$, whose role is to enforce $S_1=S_2\equiv S$ in the steady state (see Sec.~\ref{sec:examplecorrectionintwohyperspins01}). \textbf{(b)} Illustration of the directionality of the coupling between oscillators and equalizers. Solid blue (dashed magenta) line denotes that the coupling constant $\delta_{\rm eq}$ appears with the minus (plus) sign when going from the POs $A_j$ to the equalizers $Y_k$,  and with plus (minus) sign when going from the equalizers to the POs [Eqs.~\eqref{eq:dynamicsamplitueheterogeneity0101}-\eqref{eq:dynamicsamplitueheterogeneity0103}].}
\label{fig:schemeamplitudeheterogeneity1}
\end{figure}

\begin{table}[t]
\centering
\begin{tabular}{|c|c|c|}
\hline
\textbf{Coupled nodes} & \textbf{Coupling and type}\footnote{S means ``Symmetric'', and A means ``Antisymmetric''.} \\\hline
POs in different hyperspins & Linear S ($\mathbf{C}$)\\\hline
POs in the same hyperspin & Nonlinear S ($\mathbf{W}$)\\\hline
Equalizers & Decoupled \\\hline
POs and equalizers & Nonlinear A ($\mathbf{V}$)\\\hline
\end{tabular}
\caption{Type of coupling between different nodes in the hyperspin machine with amplitude heterogeneity correction (Fig.~\ref{fig:schemeamplitudeheterogeneity1}). In the hyperspin machine layer, POs belonging to different hyperspins are coupled by a symmetric (S) linear coupling $\mathbf{C}$. POs in a given hyperspin are coupled by a symmetric nonlinear coupling $\mathbf{W}$. In the equalization layer, different equalizers are not coupled between themselves; They are coupled to the POs of the hyperspin machine via an antisymmetric (A) nonlinear coupling $\mathbf{V}$ [Eqs.~\eqref{eq:dynamicsamplitueheterogeneity010302} and~\eqref{eq:dynamicsamplitueheterogeneity010303}].}
\label{table:couplingandtypeofcouplinghyperspinmachine01}
\end{table}

Figure~\ref{fig:schemeamplitudeheterogeneity1} shows the case with $D=2$ (XY hyperspins). The scheme consists of two distinct layers as in panel \textbf{(a)}: (i) The hyperspins (central layer, colored dots in the gray circles) implemented as discussed in Sec.~\ref{sec:parametricoscillators03} and Fig.~\ref{fig:sketchfixedpoints02}\textbf{b}, by nonlinearly coupling $D$ POs with amplitudes $A_1,\ldots,A_{DN}$ forming the vectors $\vec{S}_q=(A_{1+(q-1)D},\ldots,A_{qD})$, and (ii) The equalizers, with amplitudes $Y_1,\ldots,Y_M$ (top and bottom layers, yellow dots). The coupling in the composite system of PO amplitudes and equalizers is as follows (see Table~\ref{table:couplingandtypeofcouplinghyperspinmachine01}):
\begin{itemize}
\item In the hyperspin machine layer, POs $A_j$ belonging to the same hyperspin $\vec{S}_q$ are coupled by the symmetric nonlinear coupling $\mathbf{W}$ in Eq.~\eqref{eq:nonlineardynamicsparametricoscillator06}. POs belonging to different hyperspins are coupled by the linear coupling $\mathbf{C}$ in Eq.~\eqref{eq:nonlineardynamicsparametricoscillator07bis01}. We adopt $\mathbf{G}_D=\mathbbb{1}_D$ for Euclidian metric to implement the scalar product $\vec{S}_p\cdot\vec{S}_q$ as in Eq.~\eqref{eq:nonlineardynamicsparametricoscillator10};
\item In the equalization layers, different equalizers $Y_k$ are decoupled;
\item The coupling between POs $A_j$ in the hyperspin machine layer and the equalizers $Y_k$ is a nonlinear \emph{antisymmetric} coupling described by the matrix $\mathbf{V}$ as discussed below.
\end{itemize}
Similar to the error correction scheme in the coherent Ising machine~\cite{PhysRevLett.122.040607}, the equalizers evolve in time in order to drive the full dynamical system to the fixed point where all hyperspin amplitudes $S_q$ are equal, i.e., $S_q\equiv S>0$ for all $q$. At this fixed point, the equalizers $Y_k$ are stationary. For the two XY-hyperspin system in Fig.~\ref{fig:schemeamplitudeheterogeneity1}, the dynamical equations for the equalizers read
\begin{subequations}
\begin{align}
\frac{dY_1}{dt}&=\frac{\delta_{\rm eq}}{2}\left(A^2_1+A^2_2-A^2_3-A^2_4\right)Y_1 \\
\frac{dY_2}{dt}&=-\frac{\delta_{\rm eq}}{2}\left(A^2_1+A^2_2-A^2_3-A^2_4\right)Y_2 \,\, ,
\end{align}
\label{eq:dynamicsamplitueheterogeneity0101}
\end{subequations}
where $A^2_1+A^2_2\equiv S^2_1$ and $A^2_3+A^2_4\equiv S^2_2$, therefore $A^2_1+A^2_2-A^2_3-A^2_4=S^2_1-S^2_2$, and $\delta_{\rm eq}>0$ quantifies the coupling strength between the POs and the equalizers. Moreover, the equations for the PO amplitudes read
\begin{subequations}
\begin{align}
\frac{dA_1}{dt}&=\left(\frac{h_1}{4}-\frac{g}{2}\right)A_1-\frac{\beta}{2}\left(A^2_1+A^2_2\right)A_1+\frac{c}{2}A_3\nonumber\\
&-\frac{\delta_{\rm eq}}{2}\left(Y^2_1-Y^2_2\right)A_1\\
\frac{dA_2}{dt}&=\left(\frac{h_1}{4}-\frac{g}{2}\right)A_2-\frac{\beta}{2}\left(A^2_1+A^2_2\right)A_2+\frac{c}{2}A_4\nonumber\\
&-\frac{\delta_{\rm eq}}{2}\left(Y^2_1-Y^2_2\right)A_2 \,\, ,
\end{align}
\label{eq:dynamicsamplitueheterogeneity0102}
\end{subequations}
for the first hyperspin $\vec{S}_1=(A_1,A_2)$. For the second hyperspin $\vec{S}_2=(A_3,A_4)$ one has
\begin{subequations}
\begin{align}
\frac{dA_3}{dt}&=\left(\frac{h_2}{4}-\frac{g}{2}\right)A_3-\frac{\beta}{2}\left(A^2_3+A^2_4\right)A_3+\frac{c}{2}A_1\nonumber\\
&+\frac{\delta_{\rm eq}}{2}\left(Y^2_1-Y^2_2\right)A_3\\
\frac{dA_4}{dt}&=\left(\frac{h_2}{4}-\frac{g}{2}\right)A_4-\frac{\beta}{2}\left(A^2_3+A^2_4\right)A_4+\frac{c}{2}A_2\nonumber\\
&+\frac{\delta_{\rm eq}}{2}\left(Y^2_1-Y^2_2\right)A_4\,\, .
\end{align}
\label{eq:dynamicsamplitueheterogeneity0103}
\end{subequations}
As illustrated in Fig.~\ref{fig:schemeamplitudeheterogeneity1}\textbf{b}, the coupling between $Y_k$ and $A_j$ is nonlinear and directional, specifically antisymmetric: $Y_1$ couples to $A^2_1$ and $A^2_2$ with $\delta_{\rm eq}$ (solid blue connections), and it couples to $A^2_3$ and $A^2_4$ with $-\delta_{\rm eq}$ (dashed magenta connections). Similarly, $Y_2$ couples to $A^2_1$ and $A^2_2$ with $-\delta_{\rm eq}$, while it couples to $A^2_3$ and $A^2_4$ with $\delta_{\rm eq}$ [Eq.~\eqref{eq:dynamicsamplitueheterogeneity0101}]. Conversely, $A_1$ and $A_2$ couple to $Y^2_1$ with $-\delta_{\rm eq}$, and to $Y^2_2$ with $\delta_{\rm eq}$ [Eq.~\eqref{eq:dynamicsamplitueheterogeneity0102}], while $A_3$ and $A_4$ couple to $Y^2_1$ with $\delta_{\rm eq}$, and to $Y^2_2$ with $-\delta_{\rm eq}$ [Eq.~\eqref{eq:dynamicsamplitueheterogeneity0103}].

The terms in Eqs.~\eqref{eq:dynamicsamplitueheterogeneity0102} and~\eqref{eq:dynamicsamplitueheterogeneity0103} proportional to $h_1$ and $h_2$ (pump amplitudes), $g$ (intrinsic loss), $\beta$ (nonlinear saturation), and $c$ (linear coupling) compose the dynamics of the unconstrained hyperspin machine~\cite{strinati2022hyperspinmachine}. The additional terms proportional to $\delta_{\rm eq}$, jointly with the dynamics of the equalizers in Eqs.~\eqref{eq:dynamicsamplitueheterogeneity0101}, enforce $S_1=S_2$ in the steady state as follows. If during the time evolution $S_1>S_2$, then from Eqs.~\eqref{eq:dynamicsamplitueheterogeneity0101} one sees that $Y^2_1$ increases, while $Y^2_2$ decreases. This implies $Y^2_1>Y^2_2$. Accordingly, $A_1$ and $A_2$ in Eqs.~\eqref{eq:dynamicsamplitueheterogeneity0102} are subject to a negative feedback (i.e., they decrease in absolute value) by the increasing $Y^2_1-Y^2_2$ term, while $A_3$ and $A_4$ in Eqs.~\eqref{eq:dynamicsamplitueheterogeneity0102} are subject to a positive feedback (i.e., they increase in absolute value). This in turn causes $S_1$ to decrease and $S_2$ to increase. The reversed scenario is found when $S_1<S_2$. In this case, the equalization feedback is such that it causes $S_1$ to increase and $S_2$ to decrease. The mutual interaction between the POs and equalizer occurs until the condition $S_1=S_2$ is reached, where the dynamics becomes stationary.

Additional insights are provided by the fixed point of the system in Eqs.~\eqref{eq:dynamicsamplitueheterogeneity0101}-\eqref{eq:dynamicsamplitueheterogeneity0103}. The fixed points are found by equating the right-hand-side of the equations of motion to zero, and by solving for $A_j$ and $Y_k$. As we are interested in equalizing the hyperspin amplitudes $S_q$, and not the individual PO amplitudes $A_j$, the fixed points of interest are those of the dynamics of $S_1$ and $S_2$, derived from Eqs.~\eqref{eq:dynamicsamplitueheterogeneity0102} and~\eqref{eq:dynamicsamplitueheterogeneity0103} as $(1/2)dS^2_1/dt=S_1(dS_1/dt)=A_1(dA_1/dt)+A_2(dA_2/dt)$, and similarly for $S_2$:
\begin{subequations}
\begin{align}
\frac{1}{2}\frac{dS^2_1}{dt}&=\left[\frac{h_1}{4}-\frac{g}{2}-\frac{\beta}{2}S^2_1-\frac{\delta_{\rm eq}}{2}\left(Y^2_1-Y^2_2\right)\right]S^2_1\nonumber\\
&+\frac{c}{2}\,\vec{S}_1\cdot\vec{S}_2\\
\frac{1}{2}\frac{dS^2_2}{dt}&=\left[\frac{h_2}{4}-\frac{g}{2}-\frac{\beta}{2}S^2_2+\frac{\delta_{\rm eq}}{2}\left(Y^2_1-Y^2_2\right)\right]S^2_2\nonumber\\
&+\frac{c}{2}\,\vec{S}_1\cdot\vec{S}_2\,\,.
\end{align}
\label{eq:dynamicsamplitueheterogeneity03}
\end{subequations}
At the steady state, when $S_1=S_2=S>0$, the right-hand side of Eqs.~\eqref{eq:dynamicsamplitueheterogeneity03} is equal to zero, which gives ($\vec{S}_1=S\vec{\sigma}_1$ and $\vec{S}_2=S\vec{\sigma}_2$)
\begin{subequations}
\begin{align}
&\frac{h_1}{2}-g-\beta\,S^2-\delta_{\rm eq}\left(Y^2_1-Y^2_2\right)+c\,\vec{\sigma}_1\cdot\vec{\sigma}_2=0\\
&\frac{h_2}{2}-g-\beta\,S^2+\delta_{\rm eq}\left(Y^2_1-Y^2_2\right)+c\,\vec{\sigma}_1\cdot\vec{\sigma}_2=0 \,\, .
\end{align}
\label{eq:dynamicsamplitueheterogeneity04}
\end{subequations}
By taking the half sum of Eqs.~\eqref{eq:dynamicsamplitueheterogeneity04}, one obtains the fixed-point value of $S$, i.e.
\begin{equation}
S=\sqrt{\frac{1}{\beta}\left(\frac{h_1+h_2}{4}-g+c\,\vec{\sigma}_1\cdot\vec{\sigma}_2\right)} \,\, ,
\label{eq:dynamicsamplitueheterogeneity05}
\end{equation}
and by taking the half difference, one obtains the fixed-point value of $Y^2_1-Y^2_2$, i.e.
\begin{equation}
Y^2_1-Y^2_2=\frac{h_1-h_2}{4\delta_{\rm eq}} \,\, .
\label{eq:dynamicsamplitueheterogeneity06}
\end{equation}
Notice that in Eq.~\eqref{eq:dynamicsamplitueheterogeneity05} the effective loss is $g-c\,\vec{\sigma}_1\cdot\vec{\sigma}_2$, and also $-c\,\vec{\sigma}_1\cdot\vec{\sigma}_2=H_D/N$, i.e., the $D$-vector model energy in Eq.~\eqref{eq:dynamicsamplitueheterogeneity07} with $N=2$ and $D=2$ (XY energy), and coupling matrix $J_{12}=J_{21}=c$. Then the state with lowest loss is $\vec{\sigma}_1\cdot\vec{\sigma}_2=1$ for $c>0$ (ferromagnetic configuration), while $\vec{\sigma}_1\cdot\vec{\sigma}_2=-1$ for $c<0$ (antiferromagnetic configuration), i.e.
\begin{equation}
c\,\vec{\sigma}_1\cdot\vec{\sigma}_2=|c| \,\, ,
\label{eq:dynamicsamplitueheterogeneity07bis0}
\end{equation}
in Eq.~\eqref{eq:dynamicsamplitueheterogeneity05}.

\begin{figure}[t]
\centering
\includegraphics[width=8.6cm]{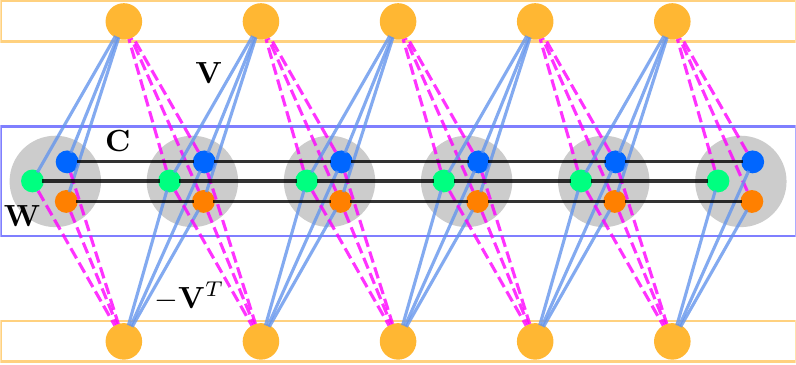}
\caption{Scheme of the multi-layer connectivity of the hyperspin machine with amplitude heterogeneity correction, for generic $N$ and $D$ (here $D=3$ is chosen for illustration purposes only). The $N$, $D$-dimensional hyperspins $\vec{S}_q$ are ordered on an open chain. Each pair of hyperspins $\vec{S}_q$ and $\vec{S}_{q+1}$ couples to a pair of equalizers as explained in Fig.~\ref{fig:schemeamplitudeheterogeneity1}, which enforce $S_q=S_{q+1}\equiv S$ (pair amplitude equalization) in the steady state. Pair amplitude equalization is enforced by the full layer of equalizers for all $q$, therefore achieving the conditions $S_q\equiv S$ for all $q$ (amplitude equalization of all hyperspin amplitudes). The three couplings (nonlinear symmetric $\mathbf{W}$ between POs forming an hyperspin, linear symmetric $\mathbf{C}$ between POs in different hyperspins, and nonlinear antisymmetric $\mathbf{V}$ between POs and equalizers) are also shown.}
\label{fig:schemeamplitudeheterogeneity2}
\end{figure}

It is worth mentioning explicitly that, when $h_1\neq h_2$ and $\delta_{\rm eq}=0$ (no amplitude equalization), the steady-state values of $S_1$ and $S_2$ are different. This is seen by equating to zero Eqs.~\eqref{eq:dynamicsamplitueheterogeneity03} without assuming equalized amplitudes, which gives the condition
\begin{equation}
\left(\beta+c\,\vec{\sigma}_1\cdot\vec{\sigma}_2\,\frac{1}{S_1S_2}\right)\left(S^2_1-S^2_2\right)=\frac{h_1-h_2}{2} \,\, .
\label{eq:dynamicsamplitueheterogeneity12}
\end{equation}
The physical interpretation of Eqs.~\eqref{eq:dynamicsamplitueheterogeneity05}, \eqref{eq:dynamicsamplitueheterogeneity06}, and~\eqref{eq:dynamicsamplitueheterogeneity12} is hence understood: When $h_1\neq h_2$ and $\delta_{\rm eq}=0$, the steady state amplitudes $S_1$ and $S_2$ are not equal, and their difference is proportional to $h_1-h_2$ [Eq.~\eqref{eq:dynamicsamplitueheterogeneity12}]. Instead, when amplitude equalization is included ($\delta_{\rm eq}>0$), the amplitude inhomogeneity in Eq.~\eqref{eq:dynamicsamplitueheterogeneity12} is ``transferred'' to the equalizers [Eq.~\eqref{eq:dynamicsamplitueheterogeneity06}], and the hyperspin amplitudes are equalized [Eq.~\eqref{eq:dynamicsamplitueheterogeneity05}].

Before continuing, two clarifications are in order. First, while in the Ising machine amplitude heterogeneity among the different parametric oscillators can lead to an improper mapping between the coupled oscillator energy and the simulated Ising Hamiltonian~\cite{PhysRevLett.122.040607}, the hyperspin machine critically requires parametric oscillator amplitudes \emph{within} a given hyperspin to be largely heterogeneous, as the simulated multidimensional vector is defined by the individual parametric oscillator components sharing the same nonlinear coupling (see Sec.~\ref{sec:parametricoscillators02}). Instead, the amplitude heterogeneity between \emph{different} hyperspins may cause the oscillator energy to differ from that of the simulated multidimensional spin vector, as in Eq.~\eqref{eq:dynamicsamplitueheterogeneity07bis0}. As such, the correction of amplitude heterogeneity in the hyperspin machine is done by requiring that the amplitudes of different hyperspin vectors are equalized in the steady state, but the amplitudes of the individual oscillators inside every hyperspins remain unconstrained. This explains the nonlinear nature of the coupling between the equalizers and the parametric oscillators in Eqs.~\eqref{eq:dynamicsamplitueheterogeneity0101} and~\eqref{eq:dynamicsamplitueheterogeneity0102}, i.e., $S^2_q$ must be equalized.

Second, the equalization scheme here proposed uses \emph{two} equalizers for a \emph{pair} of oscillators, and not one ``error-correction'' variable for each oscillator as in the case of the Ising machine~\cite{PhysRevLett.122.040607}. This fact is also a consequence of the nonlinear nature of the coupling between the equalizers and the parametric oscillators. As the equalizers appear with their squared amplitudes to the right-hand side of the equation of the oscillators [Eq.~\eqref{eq:dynamicsamplitueheterogeneity0102}], their difference should appear in order for the equalizers to reduce and augment the value of the oscillator amplitude. The equalization amplitude is here self-regulated by the dynamics and it does not require additional hyperparameters, at variance with the Ising machine.

\subsection{Equalization for general $N$ and $D$}
\label{sec:examplecorrectionintwohyperspins02}
We now move to the dynamics of the hyperspin machine with equalization layer for any $N$ and $D$, and general coupling $\mathbf{C}=\mathbf{J}_N\otimes\mathbbb{1}_D$, extending the minimal block in Fig.~\ref{fig:schemeamplitudeheterogeneity1}. For $N>2$, inhomogeneous amplitudes are usually caused by a non-homogeneous coupling elements $J_{pq}$. Therefore, in the following, we consider equal pump amplitudes for all parametric oscillators as in the original formulation of the hyperspin machine~\cite{strinati2022hyperspinmachine}.

We proceed with the system construction as follows (see Fig.~\ref{fig:schemeamplitudeheterogeneity2}). First, the $N$, $D$-dimensional hyperspins $\vec{S}_q$ are ordered on an open chain, and then for every pair of nearest-neighbour hyperspins $\vec{S}_q$ and $\vec{S}_{q+1}$ two equalizers are placed. The role of the two equalizers for each \emph{pair} of hyperspins is to enforce $S_q=S_{q+1}\equiv S$ as explained in Sec.~\ref{sec:examplecorrectionintwohyperspins01}. When pair equalization occurs for all $q$, one consequently obtains the equalization condition $S_q=S$, for all $q$.

The equations of motion for the full dynamical system in Fig.~\ref{fig:schemeamplitudeheterogeneity2} are found by generalizing Eqs.~\eqref{eq:dynamicsamplitueheterogeneity0101}-\eqref{eq:dynamicsamplitueheterogeneity0103}. For convenience, we express the dynamics in a more compact form by introducing the generalized amplitude $Z_l$ with $l=1,\ldots,DN+M$, such that $Z_l=A_l$ for $l\leq DN$, and $Z_l=Y_{l-DN}$ for $l>DN$. By inspection, one sees that the dynamics reads
\begin{eqnarray}
&&\frac{dZ_l}{dt}=-\frac{Z_l}{2}\sum_{r=1}^{DN+M}\Lambda_{lr}Z_r^2\nonumber\\
&&+\left[\left(\frac{h}{4}-\frac{g}{2}\right)Z_l+\frac{1}{2}\sum_{k=1}^{DN}C_{lk}Z_k\right]\Theta\left[DN-l\right] \,\, ,
\label{eq:dynamicsamplitueheterogeneity010301}
\end{eqnarray}
where $\Theta[n]$ is the unit step function such that $\Theta[n]=1$ for $n\geq1$ and zero otherwise. The nonlinear coupling matrix $\mathbf{\Lambda}$ is formed by the blocks
\begin{equation}
\mathbf{\Lambda}=\left(\begin{array}{c|c}\mathbf{W}&-\mathbf{V}^T\\\hline\mathbf{V}&\mathbf{0}\end{array}\right) \,\, ,
\label{eq:dynamicsamplitueheterogeneity010302}
\end{equation}
where the $M\times DN$ matrix $\mathbf{V}$ expresses the nonlinear coupling between the equalizers and POs, and it reads
\begin{equation}
\mathbf{V}=\left(\begin{array}{cccccccc}
-\vec{\delta}_{\rm eq} & \vec{\delta}_{\rm eq} & \vec{0} & \vec{0} & \cdots & \vec{0} & \vec{0}\\
\vec{\delta}_{\rm eq} & -\vec{\delta}_{\rm eq} & \vec{0} & \vec{0} & \cdots & \vec{0} & \vec{0}\\
\vec{0} & -\vec{\delta}_{\rm eq} & \vec{\delta}_{\rm eq} & \vec{0} & \cdots & \vec{0} & \vec{0}\\
\vec{0} & \vec{\delta}_{\rm eq} & -\vec{\delta}_{\rm eq} & \vec{0} & \cdots & \vec{0} & \vec{0}\\
\vdots & & & \ddots & & & \vdots\\
\vec{0} & \vec{0} & \vec{0} & \vec{0} & \cdots & -\vec{\delta}_{\rm eq} & \vec{\delta}_{\rm eq}\\
\vec{0} & \vec{0} & \vec{0} & \vec{0} & \cdots & \vec{\delta}_{\rm eq} & -\vec{\delta}_{\rm eq}
\end{array}\right) \,\, ,
\label{eq:dynamicsamplitueheterogeneity010303}
\end{equation}
where $\vec{\delta}_{\rm eq}$ and $\vec{0}$ are the $D$-dimensional row vectors with all elements equal to $\delta_{\rm eq}$ and zero, respectively.

The hyperspin heterogeneity caused by the linear coupling $\mathbf{C}$ is transferred to the equalization layer as discussed in Sec.~\ref{sec:examplecorrectionintwohyperspins01}, until the system reaches the steady state of equal amplitudes. From Eq.~\eqref{eq:dynamicsamplitueheterogeneity010301}, when the fixed point of the dynamics is reached, one has ($l\leq DN$)
\begin{equation}
-\frac{Z_l}{2}\sum_{r=1}^{DN+M}\!\!\Lambda_{lr}Z_r^2+\left(\frac{h}{4}-\frac{g}{2}\right)Z_l+\frac{1}{2}\sum_{k=1}^{DN}C_{lk}Z_k=0 \,\, .
\label{eq:amplitudehomogeneityhyperpsin18bis0}
\end{equation}
By using the form of the nonlinear coupling matrix $\mathbf{\Lambda}$ in Eq.~\eqref{eq:dynamicsamplitueheterogeneity010302}, the first summation in Eq.~\eqref{eq:amplitudehomogeneityhyperpsin18bis0} can be written by separating the $\mathbf{W}$ and the $\mathbf{V}$ terms as
\begin{eqnarray}
&&\left(\frac{h}{4}-\frac{g}{2}\right)Z_l-Z_l\frac{\beta}{2}\,\sum_{r=r_{\rm min}}^{r_{\rm min}+D-1}Z^2_r\nonumber\\
&&-\frac{Z_l}{2}\sum_{r=1}^{M}(-\mathbf{V}^T)_{lr}Z^2_{r+DN}+\frac{1}{2}\sum_{k=1}^{DN}C_{lk}Z_k=0 \,\, ,
\label{eq:amplitudehomogeneityhyperpsin18bis1}
\end{eqnarray}
where $r_{\rm min}=1+D\lfloor(l-1)/D\rfloor$. By assuming $Z_l\neq0$, one can multiply both sides of Eq.~\eqref{eq:amplitudehomogeneityhyperpsin18bis1} by $Z_l$ and sum over $l$, and obtain
\begin{eqnarray}
&&-\frac{\beta}{2}\,\sum_{l=1}^{DN}Z^2_l\sum_{r=r_{\rm min}}^{r_{\rm min}+D-1}Z^2_r-\frac{1}{2}\sum_{l=1}^{DN}\,\sum_{r=1}^{M}(-\mathbf{V}^T)_{lr}Z^2_lZ^2_{r+DN}\nonumber\\
&&+\sum_{l=1}^{DN}\left(\frac{h}{4}-\frac{g}{2}\right)Z^2_l+\frac{1}{2}\sum_{l,k=1}^{DN}C_{lk}Z_lZ_k=0 \,\, .
\label{eq:amplitudehomogeneityhyperpsin18bis2}
\end{eqnarray}
One can see that the first term in Eq.~\eqref{eq:amplitudehomogeneityhyperpsin18bis2} is
\begin{equation}
\sum_{l=1}^{DN}Z^2_l\sum_{r=r_{\rm min}}^{r_{\rm min}+D-1}Z^2_r=\sum_{q=1}^{N}S^4_q \,\, .
\label{eq:amplitudehomogeneityhyperpsin18bis3}
\end{equation}
Furthermore, by virtue of the structure of $\mathbf{V}$ in Eq.~\eqref{eq:dynamicsamplitueheterogeneity010303}, for the second term in Eq.~\eqref{eq:amplitudehomogeneityhyperpsin18bis2} one has
\begin{eqnarray}
&&\sum_{l=1}^{DN}\,\sum_{r=1}^{M}(-\mathbf{V}^T)_{lr}Z^2_lZ^2_{r+DN}\nonumber\\
&&=\delta_{\rm eq}\sum_{q=1}^{N-1}\left(Y^2_{2q-1}-Y^2_{2q}\right)\left(S^2_{q}-S^2_{q+1}\right) \,\, .
\label{eq:amplitudehomogeneityhyperpsin18bis4}
\end{eqnarray}
The linear coupling term in Eq.~\eqref{eq:amplitudehomogeneityhyperpsin18bis2}, by using the fact that $\mathbf{C}=\mathbf{J}_N\otimes\mathbbb{1}_D$, becomes~\cite{strinati2022hyperspinmachine}
\begin{equation}
\sum_{l,k=1}^{DN}C_{lk}Z_lZ_k=\sum_{p,q=1}^{N}J_{pq}\,\vec{S}_p\cdot\vec{S}_q \,\, .
\label{eq:amplitudehomogeneityhyperpsin18bis5}
\end{equation}
Finally, one can readily see that
\begin{equation}
\sum_{l=1}^{DN}Z^2_l=\sum_{q=1}^{N}S^2_q \,\, .
\label{eq:amplitudehomogeneityhyperpsin18bis501}
\end{equation}
At the steady state, where $S_q=S$ for all $q$, one has that Eq.~\eqref{eq:amplitudehomogeneityhyperpsin18bis3} equals $NS^4$, while Eq.~\eqref{eq:amplitudehomogeneityhyperpsin18bis4} yields zero. Furthermore, Eq.~\eqref{eq:amplitudehomogeneityhyperpsin18bis5} becomes $-S^2H_D$, where $H_D$ is as in Eq.~\eqref{eq:dynamicsamplitueheterogeneity07}, and Eq.~\eqref{eq:amplitudehomogeneityhyperpsin18bis501} becomes $NS^2$. Therefore, at the steady state with equalized hyperspin amplitudes, Eq.~\eqref{eq:amplitudehomogeneityhyperpsin18bis2} becomes
\begin{equation}
-\frac{\beta}{2}\,NS^4+\left(\frac{h}{4}-\frac{g}{2}\right)NS^2-\frac{1}{2}\,S^2H_{D}=0 \,\, ,
\label{eq:amplitudehomogeneityhyperpsin18bis6}
\end{equation}
from which one obtains the steady-state equalization amplitude
\begin{equation}
S=\sqrt{\frac{1}{\beta}\left(\frac{h}{2}-g-\frac{H_D}{N}\right)} \,\, .
\label{eq:amplitudehomogeneityhyperpsin18bis7}
\end{equation}
In Eq.~\eqref{eq:amplitudehomogeneityhyperpsin18bis7}, similar to what discussed in Sec.~\ref{sec:examplecorrectionintwohyperspins01}, the term $g+H_D/N$ identifies an effective loss that depends on the hyperspin energy. The minimum value of pump amplitude $h_{\rm min}$ such that $S$ in Eq.~\eqref{eq:amplitudehomogeneityhyperpsin18bis7} is real is defined by the value of $h$ that drives the lowest-loss hyperspin configuration above threshold. Since $H_D$ can take positive and negative values, $h_{\rm min}$ is given by $h_{\rm min}=2(g+E_{D,{\rm min}}/N)$, where $E_{D,{\rm min}}$ is as in Eq.~\eqref{eq:amplitudehomogeneityhyperpsin18bis7bis1}, generalizing what reported in previous work on Ising machines~\cite{PhysRevE.95.022118}.

The result in Eq.~\eqref{eq:amplitudehomogeneityhyperpsin18bis7} predicts the existence of a stationary state with equalized hyperspin amplitudes when $S^2>0$. Before validating the hyperspin machine with amplitude heterogeneity correction for large values of $N$, it is important to analyze under what conditions the steady state with equalized amplitudes $S$ in Eq.~\eqref{eq:amplitudehomogeneityhyperpsin18bis7} exists as $N$ is scaled.

\begin{figure*}[t]
\centering
\includegraphics[width=17.8cm]{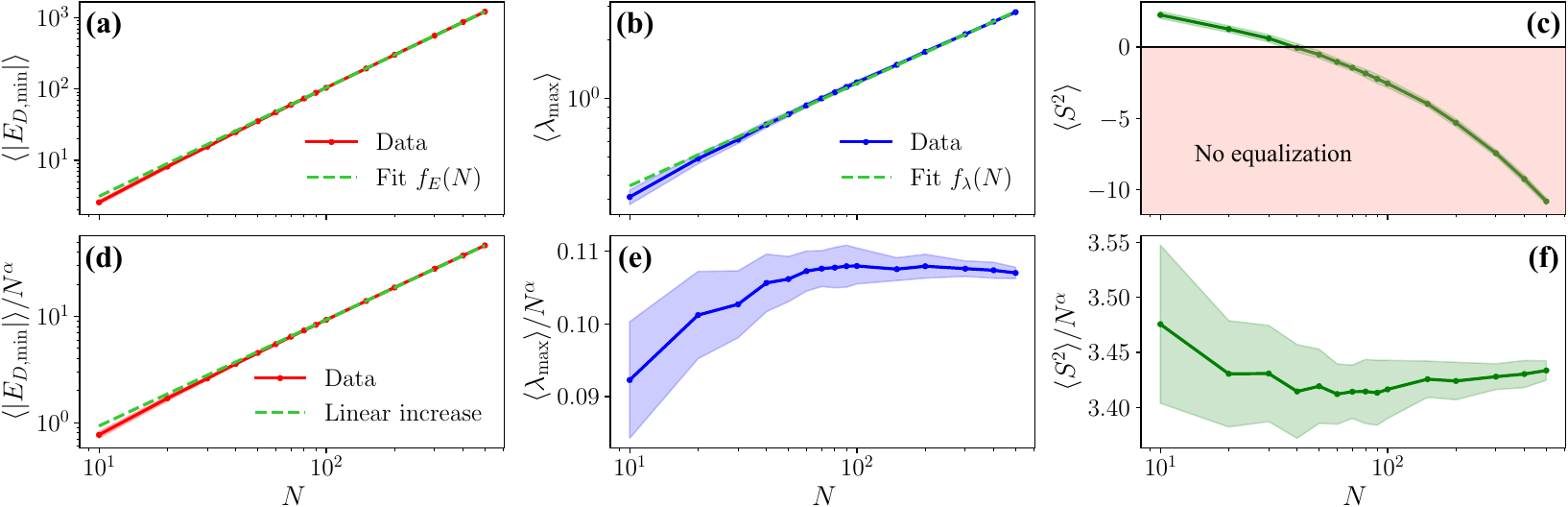}
\caption{Finite size scaling of \textbf{(a)} Average minimum energy $E_{D,{\rm min}}$ [Eq.~\eqref{eq:amplitudehomogeneityhyperpsin18bis7bis1}] in absolute value, computed by python numerical minimizer (see text), \textbf{(b)} Largest eigenvalue $\lambda_{\rm max}$ [Eq.~\eqref{eq:amplitudehomogeneityhyperpsin18bis7bis2}], and \textbf{(c)} Squared equalization amplitude $S^2$ [Eq.~\eqref{eq:amplitudehomogeneityhyperpsin18bis8}] for $\delta h=0.4$, $\beta=0.1$, and $g=1$, where the black horizontal line marks the value $S^2=0$ below which no equalized solution exists. Averages are computed over $100$ random sparse binary graphs $\mathbf{J}$, for $D=2$ and $N$ ranging from $10$ to $500$. The uncertainties (shaded areas) are quantified by the interquartile range (IQR). The graph average is denoted by $\langle\cdot\rangle$. The data in panels \textbf{(a)} and \textbf{(b)} are fitted by the power law $f_l(N)=c_l\,N^{d_l}$ where $l=E,\lambda$, respectively (green dashed line), with fitted parameters $(c_E,d_E)\simeq(0.093,1.525)$ for the energy scaling and $(c_\lambda,d_\lambda)\simeq(0.104,0.530)$ for the eigenvalue scaling (note that $d_E-d_\lambda\simeq1$). One sees from panel \textbf{(c)} that $S^2$ decreases with $N$ and becomes negative already for $N\gtrsim40$, implying that the equalization point ceases to exist as $N$ increases (red area). \textbf{(d)} Rescaled energy $E_{D,{\rm min}}/N^\alpha$ and \textbf{(e)} Maximum eigenvalue $\lambda_{\rm max}/N^\alpha$ for $\alpha=d_E-1\simeq0.525$, obtained by multiplying $\mathbf{J}$ by $1/N^\alpha$ for all $N$. This scaling makes the energy increase linearly with $N$, highlighted in the panel \textbf{(d)} by the green dashed line, and the maximum eigenvalue in panel \textbf{(e)} becomes almost independent of $N$. Accordingly, $S^2/N^\alpha$ in panel \textbf{(f)} becomes fairly constant in $N$, ensuring the existence of the equalization point for all $N$. Note the log-log scale in panels \textbf{(a)},\textbf{(b)}, and \textbf{(d)}, and the semi-log scale in panels \textbf{(c)}, \textbf{(e)}, and \textbf{(f)}.}
\label{fig:amplitudescaling01}
\end{figure*}

\section{Minimum energy and oscillation threshold scaling}
\label{sec:examplecorrectionintwohyperspins03}
In the previous section, we outlined that the steady state with equalized amplitudes depends on the pump, intrinsic loss, and the minimum hyperspin energy [Eq.~\eqref{eq:amplitudehomogeneityhyperpsin18bis7}]. These parameters depend on the chosen graph matrix $\mathbf{J}$, and therefore on $N$. If the equalized fixed point exists for a given $N$, it is not guaranteed that it still exists when scaling $N$. This is due to the fact that the minimum hyperspin energy and the oscillation threshold vary with $N$ as outlined in the following. Indeed we observe that, when increasing $N$, the convergence to the equalized amplitude state fails, due to the fact that $S^2$ becomes negative.

In this section, we show that this issue can be avoided by a suitable finite-size scaling of $\mathbf{J}$. To proceed, we express the value of pump amplitude as $h=(1+\delta h)h_{\rm th}$, where $\delta h$ is the relative deviation from the oscillation threshold value $h_{\rm th}$. In the unconstrained hyperspin dynamics [Eq.~\eqref{eq:nonlineardynamicsparametricoscillator08}], the oscillation threshold reads $h_{\rm th}=2(g-\lambda_{\rm max})$, where
\begin{equation}
\lambda_{\rm max}\coloneqq{\rm max}_j\left({\rm Re}[\lambda_j]\right) \,\, ,
\label{eq:amplitudehomogeneityhyperpsin18bis7bis2}
\end{equation}
and $\lambda_j$ are the eigenvalues of $\mathbf{C}$. We then rewrite Eq.~\eqref{eq:amplitudehomogeneityhyperpsin18bis7} for the minimum hyperspin energy  as
\begin{equation}
S=\sqrt{\frac{1}{\beta}\left[g\,\delta h-\left(1+\delta h\right)\lambda_{\rm max}+\frac{|E_{D,{\rm min}}|}{N}\right]} \,\, .
\label{eq:amplitudehomogeneityhyperpsin18bis8}
\end{equation}
One can see from Eq.~\eqref{eq:amplitudehomogeneityhyperpsin18bis8} that, for a given $\delta h$, if $\lambda_{\rm max}$ and $|E_{D,{\rm min}}|$ grow with $N$ in such a way that $S^2$ decreases with increasing $N$, the fixed point of equalized hyperspin amplitudes ceases to exist if $N$ is scaled above a threshold value. In this case, the dynamics in Eq.~\eqref{eq:dynamicsamplitueheterogeneity010301} reaches the steady state with all oscillator amplitudes $A_j$ equal to zero (trivial fixed point), while the equalizers $Y_k$ reach some nonzero value.

To ensure the functioning of the hyperspin machine with equalized amplitudes for large $N$, we therefore see that it is necessary to scale $\lambda_{\rm max}$ and $|E_{D,{\rm min}}|$ with $N$ in order to ensure that $S^2>0$ at any $N$. Since both $\lambda_{\rm max}$ and $|E_{D,{\rm min}}|$ are obtained from the graph matrix, a rescaling of $\mathbf{J}$ as
\begin{equation}
\mathbf{J}\rightarrow s\,\mathbf{J} \,\, ,
\label{eq:amplitudehomogeneityhyperpsin18bis8bis0a1}
\end{equation}
with $s>0$ rescales $\lambda_{\rm max}$ and $|E_{D,{\rm min}}|$ accordingly as $s\lambda_{\rm max}$ and $s|E_{D,{\rm min}}|$, respectively. A sufficient way to meet the condition $S^2>0$ for all $N$ is by determining the value of $s=s(N)$ from the finite-size scaling of $\lambda_{\rm max}$ and $|E_{D,{\rm min}}|$, by requiring that $S^2$ is independent of $N$. We remark that the scaling of $\mathbf{J}$ in Eq.~\eqref{eq:amplitudehomogeneityhyperpsin18bis8bis0a1} with positive $s$ does not change the ground state of the hyperspin machine but only its energy.

We proceed with this analysis by focusing on a specific class of graphs, which is random sparse binary graphs with edge density $\rho=0.4$, defined by a graph matrix $\mathbf{J}$ with random elements
\begin{equation}
J_{pq}=\left\{\begin{array}{ll}
0&\mbox{with probability $1-\rho$}\\
+J&\mbox{with probability $\rho/2$}\\
-J&\mbox{with probability $\rho/2$}
\end{array}\right. \,\, ,
\label{eq:amplitudehomogeneityhyperpsin18bis8bis01}
\end{equation}
i.e., $J_{pq}$ are random numbers extracted from a trivalued distribution with zero mean and variance $\sigma^2_J=\rho J^2$~\cite{PhysRevLett.132.017301}.

Figure~\ref{fig:amplitudescaling01} reports the statistical finite-size scaling analysis of $\lambda_{\rm max}$ and $|E_{D,{\rm min}}|$ for $D=2$, and $N$ ranging from $10$ to $500$. For each $N$, we generate $N_{\mathcal{G}}=100$ random spase graphs $\mathbf{J}^{(u)}$, where $u=1,\ldots,N_{\mathcal{G}}$ labels the graph. To compute the minimum energy, for a given value of $N$ and graph matrix $\mathbf{J}^{(u)}$ we use python \texttt{scipy.optimize.basinhopping} to minimize the XY energy defined from the polar representation of the spins $\vec{\sigma}_q=\left(\begin{smallmatrix}\cos(\varphi_q)\\\sin(\varphi_q)\end{smallmatrix}\right)$ as
\begin{equation}
H_2\equiv E_{\rm XY}(\{\varphi_q\})=-\sum_{p,q=1}^{N}J_{pq}\,\cos(\varphi_p-\varphi_q) \,\, ,
\label{eq:amplitudehomogeneityhyperpsin18bis9}
\end{equation}
with respect to the XY phases $\varphi_q$ using \texttt{L-BFGS-B} as method, giving also the gradient $\partial E_{\rm XY}/\partial\varphi_q$ to the minimizer to ease the minimization procedure. The minimization is repeated $100$ times and the minimum determined energy value out of all repetitions is selected. This value identifies the minimum energy $E^{(u)}_{D,{\rm min}}$ in Eq.~\eqref{eq:amplitudehomogeneityhyperpsin18bis7bis1} for the specific $N$ and $\mathbf{J}^{(u)}$.

Top panels \textbf{(a)}-\textbf{(c)} of Fig.~\ref{fig:amplitudescaling01} show respectively $\langle\lambda_{\rm max}\rangle$, $\langle|E_{D,{\rm min}}|\rangle$, and $\langle S^2\rangle$, where
\begin{equation}
\left\langle\left|E_{D,{\rm min}}\right|\right\rangle\coloneqq\frac{1}{N_{\mathcal{G}}}\sum_{u=1}^{N_{\mathcal{G}}}\left|E^{(u)}_{D,{\rm min}}\right| \,\, ,
\label{eq:amplitudehomogeneityhyperpsin18bis10}
\end{equation}
denotes the graph average of the minimum XY energy $|E^{(u)}_{D,{\rm min}}|$ from the selected $\mathbf{J}^{(u)}$, and similarly for $\langle\lambda_{\rm max}\rangle$ and $\langle S^2\rangle$. We see from our numerical data that the minimum energy in panel \textbf{(a)} and maximal eigenvalue in panel \textbf{(b)} are well represented for sufficiently large $N$ by a power law of the form $f_E(N)=c_EN^{d_E}$ and $f_\lambda(N)=c_\lambda N^{d_\lambda}$, respectively (dashed lines in the plots), with coefficients $c_E$ and $c_\lambda$, and exponents $d_E$ and $d_\lambda$, obtained by fitting the numerical data.

In particular, we find that $d_\lambda\simeq0.53$. This value is consistent with scaling exponent $1/2$ of the largest eigenvalue of random matrices predicted in the limit $N\rightarrow\infty$ by Bai-Yin’s law~\cite{baiyin1988}. Moreover, for the minimum energy scaling, we find $d_E\simeq1.525$, which agrees with the expected scaling exponent $3/2$ of the energy for random graphs with fixed variance $\sigma^2_J$ as in Eq.~\eqref{eq:amplitudehomogeneityhyperpsin18bis8bis01}. Indeed, the energy for random graph is expected to be extensive (i.e., scale as $N$) when the variance is rescaled as $\sigma^2_J/N$~\cite{nishimori2001statistical}. This means that $\mathbf{J}$ in Eq.~\eqref{eq:amplitudehomogeneityhyperpsin18bis8bis01} should be rescaled as $\mathbf{J}/\sqrt{N}$, and accordingly $E_{D,{\rm min}}$ is rescaled as $E_{D,{\rm min}}/\sqrt{N}$, in order for the energy to be extensive. We ascribe fact that we fit values of $d_\lambda$ and $d_E$ which slightly differ from the expected values both to finite-size effects and to the limited precision of our statistics because of the limited number of graphs $N_{\mathcal{G}}$.

\begin{figure}[t]
\centering
\includegraphics[width=8.5cm]{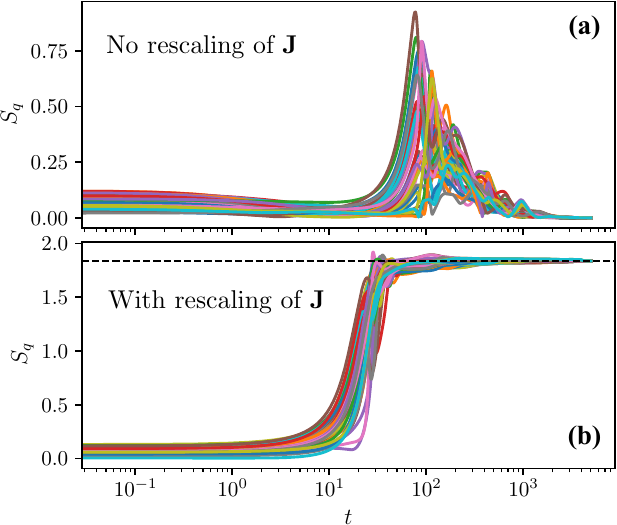}
\caption{Hyperspin amplitude time evolution $S_q(t)$ obtained from the dynamics in Eq.~\eqref{eq:dynamicsamplitueheterogeneity010301} for a random sparse graph with $N=40$ with \textbf{(a)} No rescaling of $\mathbf{J}^{(u)}$, and \textbf{(b)} Rescaling of $\mathbf{J}^{(u)}$ by $s=1/N^\alpha$ (see Sec.~\ref{sec:examplecorrectionintwohyperspins03}), numerically integrated by a second-order Runge-Kutta method with time step $0.05$ and parameters $\delta h=0.4$, $\beta=0.1$, $g=1$, and $\delta_{\rm eq}=0.4$ (see Fig.~\ref{fig:amplitudescaling01}). Without rescaling, the steady state with equalized amplitudes does not exist. The hyperspin amplitudes after a nontrivial dynamics converge to the trivial state where they are all zero. Instead, the rescaling ensures the existence of the equalized steady state, where all $S_q$ converge to $S$ in Eqs.~\eqref{eq:amplitudehomogeneityhyperpsin18bis7} and~\eqref{eq:amplitudehomogeneityhyperpsin18bis8}, marked by the horizontal black dashed line.}
\label{fig:amplitudescaling02}
\end{figure}

The analysis shows that the absolute value of the minimum energy and maximal eigenvalue increase with $N$ with scaling exponent satisfying $d_E-d_\lambda\simeq 1$. Accordingly, the average squared equalization amplitude in panel \textbf{(c)}, which from Eq.~\eqref{eq:amplitudehomogeneityhyperpsin18bis8} reads
\begin{equation}
\left\langle S^2\right\rangle\simeq\frac{g\,\delta h-\left(1+\delta h\right)c_\lambda N^{d_\lambda}+c_EN^{d_E-1}}{\beta} \,\, ,
\label{eq:amplitudehomogeneityhyperpsin18bis11}
\end{equation}
decreases with $N$, eventually becoming negative as $N$ is increased above a certain value, signifying that no equalization beyond that value of $N$ is possible.

As mentioned before, this issue is mitigated by rescaling all matrices $\mathbf{J}^{(u)}$ by $s>0$ to ensure that $S^2$ remains positive at all $N$. In our case, the result $d_E-d_\lambda\simeq 1$ allows to write Eq.~\eqref{eq:amplitudehomogeneityhyperpsin18bis11} as
\begin{equation}
\left\langle S^2\right\rangle\simeq\frac{g\,\delta h-\left[\left(1+\delta h\right)c_\lambda-c_E\right]N^{d_E-1}}{\beta} \,\, .
\label{eq:amplitudehomogeneityhyperpsin18bis12}
\end{equation}
Therefore, by choosing $s=1/N^{d_E-1}$, the coefficients $c_\lambda$ and $c_E$ are accordingly multiplied by $s$, and then the dependence on $N$ in Eq.~\eqref{eq:amplitudehomogeneityhyperpsin18bis12} is compensated. This is shown in bottom panels \textbf{(d)}-\textbf{(f)} of Fig.~\ref{fig:amplitudescaling01}, where the quantity in the corresponding top panel is shown multiplied by $s$. As evident, for sufficiently large $N$, the rescaled minimum energy varies linearly with $N$, while the maximum eigenvalue becomes almost independent of $N$. This result is consistent with the expected scalings for random graphs discussed before. Accordingly, the squared equalized amplitude becomes fairly independent of $N$, pinpointing the existence of the steady state with equal hyperspin amplitudes for all $N$. We stress that the correct rescaling of $\mathbf{J}$ is remarkably crucial for large $N$, where the hyperspin machine with unscaled $\mathbf{J}$ fails to converge to the equalized solution.

Figure~\ref{fig:amplitudescaling02} exemplifies the dynamics of the hyperspin amplitudes with unscaled and rescaled $\mathbf{J}^{(u)}$, respectively in panel \textbf{(a)} and \textbf{(b)}, for a graph matrix with $N=40$, which is the value of $N$ above which $\langle S^2\rangle$ in Fig.~\ref{fig:amplitudescaling01}\textbf{c} becomes negative. The hyperspin amplitudes $S_q$ when no rescaling of $\mathbf{J}^{(u)}$ is performed converge in the long-time limit all to zero, because when the fixed point with equalized amplitudes does not exist all oscillator amplitudes converge to $A_j=0$, for all $j$ (trivial state). In this case, the equalized hyperspin machine fails to operate. Instead, when $\mathbf{J}^{(u)}$ is rescaled as explained before, the hyperspin amplitudes converge to the fixed point with equal amplitudes $S_q=S$ for all $q$, where $S$ is given in Eqs.~\eqref{eq:amplitudehomogeneityhyperpsin18bis7} and~\eqref{eq:amplitudehomogeneityhyperpsin18bis8}.

\section{Numerical results}
\label{sec:numericalretuls1}
In this section, we present our numerical results on the equalized hyperspin machine. The goal is to compare the energy reached by the hyperspin machine with and without amplitude equalization on large-scale systems, and demonstrate that amplitude equalization allows to approach the ground state of $H_D$ with orders of magnitude increased accuracy.

\subsection{Hyperspin machine by nonlinear maps}
\label{sec:numericalretuls2}
To carry on our numerical analysis, we simulate the time evolution by resorting to the formalism of nonlinear maps~\cite{strogatz2007nonlinear}, which allows a significantly more efficient numerical simulation, as well as connecting our analytical model to an experimentally relevant setup. Specifically, we extend the hybrid electro-optical hyperspin machine model discussed in Ref.~\cite{PhysRevLett.132.017301}, which we now recall for the sake of completeness.

\begin{figure}[t]
\centering
\includegraphics[width=8.5cm]{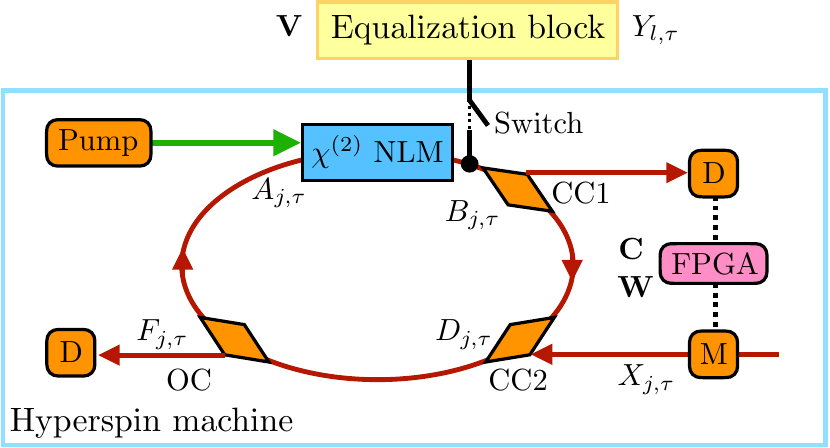}
\caption{Scheme of the nonlinear map of the electro-optical coherent hyperspin machine of Ref.~\cite{PhysRevLett.132.017301}, where amplitude heterogeneity correction is implemented as an additional block modulating the oscillator amplitudes after the nonlinear medium. The connection of the hyperspin machine to the equalization block can be controlled during the dynamics via a switch. The blocks are: Second-order nonlinear medium ($\chi^{(2)}$ NLM) driven by a pump, coupling couplers (CC1 and CC2), detector (D), field-programmable-gate-array (FPGA), modulator (M), and output coupler (OC), each one implementing a specific input-output relation on the PO fields (see Sec.~\ref{sec:numericalretuls1}).}
\label{fig:amplitudeheterogeneitycorrectionscheme1}
\end{figure}

Figure~\ref{fig:amplitudeheterogeneitycorrectionscheme1} shows the map implementation of the hyperspin machine with equalization of Fig.~\ref{fig:schemeamplitudeheterogeneity2}. The scheme consists of two distinct macro blocks: The field-programmable-gate-array (FPGA)-based hyperspin machine~\cite{PhysRevLett.132.017301}, inside the blue square, and the equalization block (yellow square), whose connection to the hyperspin machine is controlled by a switch.

The continuous-time dynamics in Eq.~\eqref{eq:dynamicsamplitueheterogeneity010301} is implemented by the nonlinear map by discretizing the time evolution as $\tau=n\,\tau_{\rm rt}$, where $\tau_{\rm rt}$ denotes the cavity round trip time, and $n$ is a positive integer. The real PO amplitudes $A_{j,\tau}$ are then represented by two labels: $j=1,\ldots,DN$ labelling the PO, and $\tau=1,\ldots,N_{\rm rt}$ labelling the round trip, where $N_{\rm rt}$ is the total round trip number. At each round trip, the main dynamical processes (amplification, coupling, intrinsic loss, and measurement) are simulated as different functional blocks, each one with a specific input-output relation, operating on the PO fields $A_{j,\tau}$ at distinct position inside the cavity.

Specifically, the hyperspin machine macro block in Fig.~\ref{fig:amplitudeheterogeneitycorrectionscheme1} implements the main dynamical processes as follows (see Refs.~\cite{PhysRevLett.126.143901,PhysRevLett.132.017301} for a detailed discussion):

\textit{(i) Parametric amplification}: A second-order nonlinear crystal ($\chi^{(2)}$ NLM) of length $L$ and nonlinear constant $\kappa$, pumped by a laser beam with uniform amplitude $h$ (i.e., independent of $j$) amplifies the PO fields $A_{j,\tau}$, which are input to the nonlinear medium together with the pump. The $\chi^{(2)}$ NLM block takes $A_{j,\tau}$ as input, and outputs
\begin{equation}
B_{j,\tau}=K_j(A_{j,\tau},h,\kappa L)\,A_{j,\tau} \,\, ,
\label{eq:nonlinearmap1}
\end{equation}
where $K_j(A_{j,\tau},h,\kappa L)$ the amplification factor computed from the solution of the second-order nonlinear wave equations~\cite{boyd2008nonlinear} under the assumptions that all fields are real and the pump field is not fully depleted inside the nonlinear medium~\cite{PhysRevLett.132.017301}:
\begin{equation}
K_j(A_{j,\tau},h,\kappa L)\coloneqq\frac{\sqrt{I_{j,\tau}}}{|A_{j,\tau}|}\,{\rm sech}\,\left(\kappa L\sqrt{I_{j,\tau}}-\xi_0\right) \,\, ,
\label{eq:nonlinearmap2}
\end{equation}
with $\xi_0=\log[(\sqrt{I_{j,\tau}}+h)/|A_{j,\tau}|]$ and $I_{j,\tau}=\sqrt{A^2_{j,\tau}+h^2}$.

\textit{(ii) Coupling}: After the nonlinear medium, both the linear $\mathbf{C}$ and nonlinear coupling $\mathbf{W}$ are implemented by the coupling device acting as follows. A coupler (CC1) extracts a fraction $\sqrt{b}$ of the $B_{j,\tau}$ fields, while the remaining part $\sqrt{a}B_{j,\tau}$ with $a+b=1$ remains inside the cavity. Then, a detector (D) transforms $\sqrt{b}B_{j,\tau}$ into digital signal and sends it to a FPGA~\cite{Marandi2016,PhysRevApplied.13.054059} preprogrammed with both the $\mathbf{C}$ and the $\mathbf{W}$ matrix. The FPGA controls a modulator (M), which reconverts the digital signal into optical field, overall transforming the input field $\sqrt{b}B_{j,\tau}$ into
\begin{equation}
X_{j,\tau}=\sqrt{b}\left(\sum_{k=1}^{DN}C_{jk}B_{k,\tau}-bB_{j,\tau}\sum_{r=1}^{DN}W_{jr}B^2_{r,\tau}\right) \,\, .
\label{eq:nonlinearmap3}
\end{equation}
A second coupler CC2, identical to CC1, recombines the $X_{j,\tau}$ with the fraction of the $B_{j,\tau}$ left inside the cavity, overall yielding the field $D_{j,\tau}=aB_{j,\tau}+\sqrt{b}X_{j,\tau}$. Therefore, the overall coupled PO field reads
\begin{equation}
D_{j,\tau}=\sum_{k=1}^{DN}Q_{jk}B_{k,\tau}-b^2B_{j,\tau}\sum_{r=1}^{DN}W_{jr}B^2_{r,\tau} \,\, .
\label{eq:nonlinearmap4}
\end{equation}
where $\mathbf{Q}=a\mathbbb{1}_{DN}+b\mathbf{C}$ is the $DN\times DN$ effective coupling matrix including self interaction ($a\mathbbb{1}_{DN}$) and linear coupling ($b\mathbf{C}$).

\textit{(iii) Measurement and intrinsic loss}: The cavity intrinsic losses are phenomenologically described at the level of the output coupler (OC) that extracts a fraction $0\leq d<1$ of the coupled PO fields $D_{j,\tau}$, and sends the measured field $F_{j,\tau}=dD_{j,\tau}$ to the detector. The remaining field $(1-d)D_{j,\tau}$ remains inside the cavity, and it identifies the input field to the nonlinear medium at the next round trip.

\begin{figure*}[t]
\centering
\includegraphics[width=17.8cm]{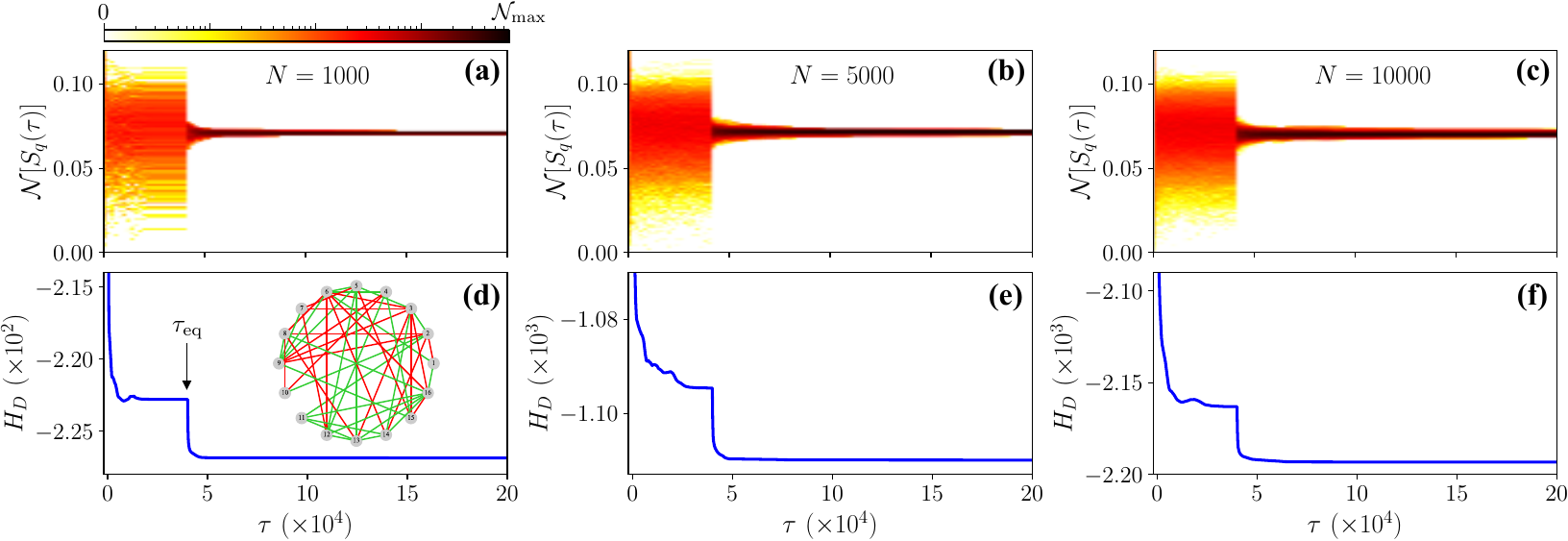}
\caption{\textbf{(a)}-\textbf{(c)} Colormap of the histogram $\mathcal{N}[S_q(\tau)]$ of the hyperspin amplitudes $S_q(\tau)$ and \textbf{(d)}-\textbf{(f)} Hyperspin energy $H_D(\tau)$ from Eq.~\eqref{eq:dynamicsamplitueheterogeneity07} with $\vec{\sigma}_q=\vec{S}_q/S_q$ during the round trips $\tau$ for \textbf{(a)},\textbf{(d)} $N=1000$, \textbf{(b)},\textbf{(e)} $N=5000$, and \textbf{(c)},\textbf{(f)} $N=10000$ as in the figure labels. Colormap coding: White for $\mathcal{N}=0$ up to black for some maximum value $\mathcal{N}=\mathcal{N}_{\rm max}$ as in the colorbar (note the logarithmic scale). For all data $D=2$ (XY hyperspins) is simulated using the scheme in Fig.~\ref{fig:amplitudeheterogeneitycorrectionscheme1} (see Sec.~\ref{sec:numericalretuls2}) on random binary sparse graph $\mathbf{J}$ with $40\%$ edge density [Eq.~\eqref{eq:amplitudehomogeneityhyperpsin18bis8bis01}], sketched in panel \textbf{(d)} with $N=16$ for illustration purposes. Green and red lines denote positive and negative coupling $J_{pq}$, respectively. In all cases, the hyperspin machine is connected to the equalization layer from round trip $\tau_{\rm eq}=4\times10^4$, marked by the black arrow in panel \textbf{(d)}. Thus, for $\tau\leq\tau_{\rm eq}$ the unconstrained hyperspin machine in Eq.~\eqref{eq:nonlinearmap5} is simulated, showing largely heterogeneous amplitudes~\cite{strinati2022hyperspinmachine,PhysRevLett.132.017301}, while for $\tau>\tau_{\rm eq}$ the equalization layer implementing the map in Eq.~\eqref{eq:nonlinearmap9} quickly induces the hyperspin amplitudes to converge to a homogeneous value $S_q=S$ for all $q$ [panels \textbf{(a)}-\textbf{(c)}]. The reduction of amplitude heterogeneity in turn significantly reduces the energy value compared to that reached by the unconstrained hyperspin machine [panels \textbf{(d)}-\textbf{(f)}] due to the proper mapping of the hyperspin energy onto the $D$-vector model energy~\cite{strinati2022hyperspinmachine}. The fact that the hyperspin amplitudes at larger $N$ appear more heterogeneous also for $\tau>\tau_{\rm eq}$ is ascribed to the increased relaxation time to the equalized fixed point for increasing $N$.}
\label{fig:nonlinarmaphistogramdynamics01}
\end{figure*}

Overall, by combining Eqs.~\eqref{eq:nonlinearmap1}-\eqref{eq:nonlinearmap4}, the nonlinear map of the hyperspin machine with no amplitude equalization, linking the PO amplitudes at round trip $\tau+1$ to those at round trip $\tau$ is then~\cite{PhysRevLett.132.017301}
\begin{equation}
A_{j,\tau+1}=(1-d)\,D_{j,\tau}(\{B_{j,\tau}\}) \,\, ,
\label{eq:nonlinearmap5}
\end{equation}
where the dependence of the $D_{j,\tau}$ fields on $B_{j,\tau}$ is explicit. The equalization block is included in the nonlinear map by discretizing the time variable in Eqs.~\eqref{eq:dynamicsamplitueheterogeneity010301}, in particular considering the terms describing the coupling between the PO amplitudes $A_j$ and the equalizers $Y_k$. In continuous time, these coupling terms read ($\mathbf{V}^T_{jk}=V_{kj}$)
\begin{equation}
\frac{dY_k}{dt}=-\frac{Y_k}{2}\sum_{j=1}^{DN}V_{kj}A^2_{j} \qquad \frac{dA_j}{dt}=\frac{A_j}{2}\sum_{k=1}^{M}V_{kj}Y^2_{k} \,\, .
\label{eq:nonlinearmap6}
\end{equation}
Here, we place the equalization block after the nonlinear medium as in Fig.~\ref{fig:amplitudeheterogeneitycorrectionscheme1}, and denote the equalizers as $Y_{k,\tau}$ where $k=1,\ldots,M$. Therefore, in the map dynamics, the PO field in Eq.~\eqref{eq:nonlinearmap6} is actually $B_{j,\tau}$, which identifies the input fields to the equalization block. These fields are transformed as (the prefactor $1/2$ is omitted)
\begin{equation}
B^{{\rm (eq)}}_{j,\tau}=\left(1+\sum_{k=1}^{M}V_{kj}Y^2_{k,\tau}\right)B_{j,\tau} \,\, ,
\label{eq:nonlinearmap7}
\end{equation}
where the superscript ``eq'' stands for ``equalized''. Similarly, the equalizers are updated as
\begin{equation}
Y_{k,\tau+1}=\left(1-\sum_{j=1}^{DN}V_{kj}B^2_{j,\tau}\right)Y_{k,\tau} \,\, .
\label{eq:nonlinearmap8}
\end{equation}
Notice that Eq.~\eqref{eq:nonlinearmap8} defines the equalizers at the next round trip because these variables are used only to equalize the PO fields, and they influence the dynamics only when the equalization block is encountered. Instead, the PO fields are further transformed by the map blocks following Eq.~\eqref{eq:nonlinearmap7}, with the difference that the coupled field $D_{j,\tau}$ are now defined in terms of $B^{{\rm (eq)}}_{j,\tau}$ instead of $B_{j,\tau}$. Therefore, the nonlinear map for the equalized hyperspin machine is
\begin{equation}
A_{j,\tau+1}=(1-d)\,D_{j,\tau}(\{B^{{\rm (eq)}}_{j,\tau}\}) \,\, .
\label{eq:nonlinearmap9}
\end{equation}
When the fixed point with equalized PO amplitudes exists (recall Sec.~\ref{sec:examplecorrectionintwohyperspins02}), the nonlinear map in Eq.~\eqref{eq:nonlinearmap9} is simulated for a sufficiently large number of round trips $N_{\rm rt}$ until the system reaches the equalized stationary state. In the following, to numerically simulate the maps in Eqs.~\eqref{eq:nonlinearmap5} and~\eqref{eq:nonlinearmap9}, we use an extended version of the parallelized C-language code developed in Ref.~\cite{PhysRevLett.132.017301}, here modified to simulate also amplitude equalization.

\subsection{Dynamics of PO amplitudes and energy}
\label{sec:dynamicsofamplitudeandenergy1}
Top panels of Fig.~\ref{fig:nonlinarmaphistogramdynamics01} show the numerically obtained data of the histogram $\mathcal{N}$ of the hyperspin amplitudes time evolution $S_q(\tau)$, computed from the PO fields $A_{j,\tau}$ as $S_q(\tau)=\sqrt{\sum_{\mu=1}^{D}A^2_{\mu+(q-1)D,\tau}}$ obtained by simulating the nonlinear map  discussed in Sec.~\ref{sec:numericalretuls2} for $N_{\rm rt}=2\times10^5$ round trips. The PO and equalizers fields are initialized at $\tau=0$ as random numbers from a uniform distribution taking both positive and negative values. Bottom panels show the corresponding hyperspin energy $H_D(\tau)$ during the round trips, computed from Eq.~\eqref{eq:dynamicsamplitueheterogeneity07} with $\vec{\sigma}_q=\vec{S}_q/S_q$.

Specifically, to highlight the different behaviour of the system with and without amplitude heterogeneity correction, we simulate the dynamics without amplitude heterogeneity correction in Eq.~\eqref{eq:nonlinearmap5} for an initial time $\tau\leq\tau_{\rm eq}$ with $\tau_{\rm eq}=4\times10^4$. Then, for $\tau>\tau_{\rm eq}$, amplitude heterogeneity correction is performed by simulating the map in Eq.~\eqref{eq:nonlinearmap9}. Pictorially, we close the switch in Fig.~\ref{fig:amplitudeheterogeneitycorrectionscheme1} at round trip $\tau=\tau_{\rm eq}$, connecting the hyperspin machine to the equalization block.

We consider three large values of $N$, which are $N=1000$ for panels \textbf{(a)},\textbf{(d)}, $N=5000$ for panels \textbf{(d)},\textbf{(e)}, and $N=10000$ for panels \textbf{(c)},\textbf{(f)}, as in the figure labels. For each $N$, the coupling matrix is $\mathbf{C}=\mathbf{J}\otimes\mathbbb{1}_D$ with $D=2$ and $\mathbf{J}$ represented by a $N\times N$ random sparse matrix of the same kind discussed in Sec.~\ref{sec:examplecorrectionintwohyperspins03}, by rescaling $\mathbf{J}$ as in Eq.~\eqref{eq:amplitudehomogeneityhyperpsin18bis8bis0a1} by
\begin{equation}
\mathbf{J}\rightarrow s\,\mathbf{J} \quad\mbox{with}\quad s=\frac{J_0}{N^\alpha} \,\, ,
\label{eq:nonlinearmap10}
\end{equation}
where $\alpha=0.525$ (see Fig.~\ref{fig:amplitudescaling01}) and $J_0=20^\alpha/2$ is chosen by convention such that $s=1/2$ for $N=20$.

As evident from panels \textbf{(a)}-\textbf{(c)} of Fig.~\ref{fig:nonlinarmaphistogramdynamics01}, when the hyperspin machine is connected to the equalization block, the hyperspin amplitudes quickly turn from being significantly heterogeneous, quantified by an histogram that spreads over a large range of values, to converging to a very narrow histogram. This fact indicates the sharp decrease of amplitude heterogeneity in the system.

After such an initial rapid heterogeneity reduction around $\tau=\tau_{\rm eq}$, we observe that the hyperspin amplitudes relax to the equalized state over an increasing time scale for larger $N$, which is ascribed not only to the increased $N$ but also to the reduced coupling constant due to the rescaling in Eq.~\eqref{eq:nonlinearmap10}. This is seen by comparing the width of the histograms $\mathcal{N}$ for large $\tau$ in the three cases: While for $N=1000$ the histogram is well peaked around a single value of $S_q$, for $N=5000$ and especially $N=10000$ the histogram shows a broader distribution, meaning that the actual reach of the equalized state occurs for $\tau$ larger than the simulation time $N_{\rm rt}$ used in this figure.

Together with the hyperspin amplitude dynamics, a deeper insight on the equalized hyperspin machine dynamics is provided by the hyperspin energy time evolution in panels \textbf{(d)}-\textbf{(f)} of Fig.~\ref{fig:nonlinarmaphistogramdynamics01}. For $\tau\leq\tau_{\rm eq}$, where the unconstrained hyperspin machine is simulated, there is an initial transient where $H_D$ decreases in time converging to a first stationary value with heterogeneous amplitudes. Such an energy value, albeit it can be very close to the actual energy minimum of the simulated $D$-vector model, corresponds in general to a local minimum of $H_D$, especially for hard graphs~\cite{strinati2022hyperspinmachine} (see also Fig.~\ref{fig:sketchfixedpoints02}\textbf{d}).

When the hyperspin machine is connected to the equalization block at $\tau>\tau_{\rm eq}$, the energy value is quickly reduced, reaching a considerably lower energy value with respect to the first one found at $\tau\leq\tau_{\rm eq}$. The correlation between the data in top and bottom panels of Fig.~\ref{fig:nonlinarmaphistogramdynamics01} is manifest: The reduction of amplitude heterogeneity between the hyperspin amplitudes $S_q$ reduces the value of the hyperspin energy $H_D$ accordingly. The results obtained in this section validate the discussion in Ref.~\cite{strinati2022hyperspinmachine}, providing first clear evidence that the reduction of amplitude heterogeneity in the hyperspin machine makes the system behave as a gradient descent minimizing a Lyapunov function that is closer to $H_D$ the more the hyperpsin amplitudes $S_q$ are equalized.

\subsection{Dependence on the pump amplitude}
\label{sec:dynamicsofamplitudeandenergy2}
\begin{figure}[t]
\centering
\includegraphics[width=8.4cm]{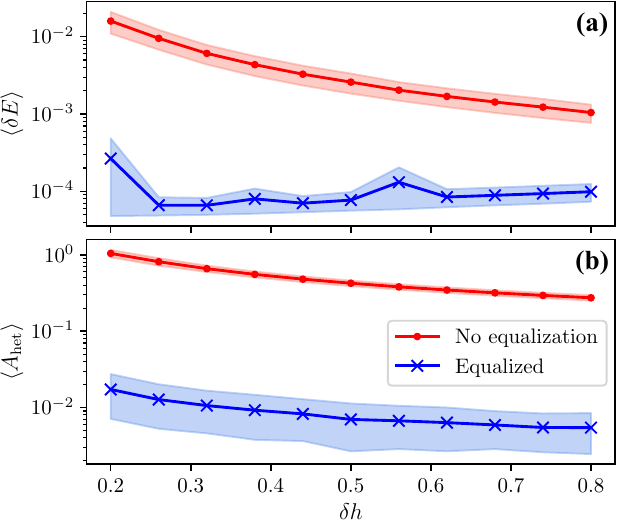}
\caption{Graph average of \textbf{(a)} Relative energy deviation $\delta E$ [Eq.~\eqref{eq:nonlinearmap11}] and \textbf{(b)} Amplitude heterogeneity degree $A_{\rm het}$ [Eq.~\eqref{eq:nonlinearmap12}] shown in semi-logarithmic scale as a function of pump relative deviation from threshold $\delta h$. Data are averaged for each pump value over $50$ random graphs. The graph average is denoted by $\langle\cdot\rangle$. Data are shown for $N=100$ and $D=2$ both without and with amplitude equalization [red and blue connected points from Eqs.~\eqref{eq:nonlinearmap5} and~\eqref{eq:nonlinearmap9}, respectively] with uncertainty quantified by the IQR (shaded areas).}
\label{fig:numericalresults1}
\end{figure}

We now analyze the ability of the equalized hyperspin machine in Eq.~\eqref{eq:nonlinearmap9} to act as a minimizer of the $D$-vector model energy $H_D$, comparing its performance with that of the hyperspin without amplitude equalization in Eq.~\eqref{eq:nonlinearmap5}. To do so, we focus on two quantities:
\begin{itemize}
\item Relative energy deviation
\begin{equation}
\delta E\coloneqq\frac{H_D-E_{D,{\rm min}}}{|E_{D,{\rm min}}|} \,\, ,
\label{eq:nonlinearmap11}
\end{equation}
quantifying the relative difference between the hyperspin energy $H_D$ in Eq.~\eqref{eq:dynamicsamplitueheterogeneity07} obtained from the hyperspin machine, and the minimum energy $E_{D,{\rm min}}$ in Eq.~\eqref{eq:amplitudehomogeneityhyperpsin18bis7bis1} computed by resorting to the python numerical minimizer as explained in Sec.~\ref{sec:examplecorrectionintwohyperspins03};
\item Heterogeneity degree
\begin{equation}
A_{\rm het}\coloneqq\frac{S_{\rm max}-S_{\rm min}}{\overline{S}} \,\, ,
\label{eq:nonlinearmap12}
\end{equation}
quantifying amplitude heterogeneity, where $S_{\rm max}=\max_q\{S_q\}$ and $S_{\rm min}=\min_q\{S_q\}$ are the maximum and minimum hyperspin amplitudes, and $\overline{S}=N^{-1}\sum_{q=1}^{N}S_q$ is the average amplitude.
\end{itemize}
Figure~\ref{fig:numericalresults1} reports a prototype result of the statistical analysis of the relative energy deviation and heterogeneity degree for $N=100$, as a function of pump amplitude relative deviation from threshold $\delta h$. The data are shown as graph averages defined similarly to Eq.~\eqref{eq:amplitudehomogeneityhyperpsin18bis10} as
\begin{equation}
\langle\delta E\rangle=\frac{2}{N_{\mathcal{G}}}\sum_{u=1}^{N_{\mathcal{G}}/2}\delta E^{(u)} \quad \langle A_{\rm het}\rangle=\frac{2}{N_{\mathcal{G}}}\sum_{u=1}^{N_{\mathcal{G}}/2}A_{\rm het}^{(u)} \,\, ,
\label{eq:nonlinearmap13}
\end{equation}
averaging over $N_{\mathcal{G}}/2=50$ random graphs picked from the set of graphs defined in Sec.~\ref{sec:examplecorrectionintwohyperspins03}, with $\mathbf{J}$ rescaling as in Eq.~\eqref{eq:nonlinearmap10}. We scan $11$ equally spaced values of the pump relative deviation from $\delta h=0.2$ to $\delta h=0.8$. For each $\delta h$ and graph $u$, we repeat the map simulation $20$ times, both without and with amplitude equalization [Eqs.~\eqref{eq:nonlinearmap5} and~\eqref{eq:nonlinearmap9}], up to $N_{\rm rt}=4\times10^4$. The choice of these numerical parameters is done to reduce the probability that the hyperspin machine converges to a local minimum~\cite{strinati2022hyperspinmachine} while keeping a reasonable numerical complexity.

\begin{figure}[t]
\centering
\includegraphics[width=8.4cm]{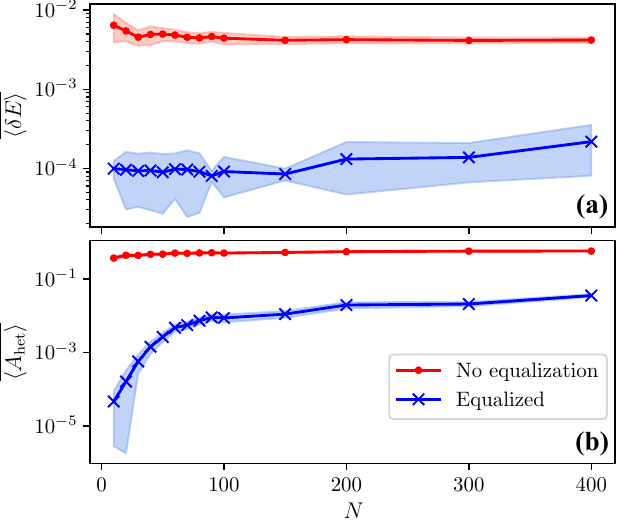}
\caption{Pump and graph average of \textbf{(a)} Relative energy deviation $\delta E$ and \textbf{(b)} Amplitude heterogeneity degree $A_{\rm het}$ shown in semi-logarithmic scale as a function $N$, obtained by averaging over pump the graph averaged data (shown in Fig.~\ref{fig:numericalresults1} specifically for $N=100$). The pump and graph average is denoted by $\overline{\langle\cdot\rangle}$. Data are shown for $D=2$ both without and with amplitude equalization (red and blue connected points, respectively) with uncertainties (shaded areas).}
\label{fig:numericalresults2}
\end{figure}

Data from Eq.~\eqref{eq:nonlinearmap5} are shown as red points and labelled as ``No equalization'', while those from Eq.~\eqref{eq:nonlinearmap9} are shown as blue crosses and labelled as ``Equalized''. The statistical uncertainty is quantified by the interquartile range (IQR) and shown as shaded areas with corresponding color of the data markers.

The results from this analysis point out two remarkable findings: First, from panel \textbf{(a)} one sees that the equalized hyperspin machine finds on average values of the relative variation that are between one and two orders of magnitude lower than those from the hyperspin machine with no equalization. Second, the relative energy deviation is significantly less sensitive to the pump amplitude value (note that the data are shown in semi-logarithmic scale). The considerable reduction of the relative energy deviation when equalization is simulated means that the equalized hyperspin machine finds with remarkably enhanced precision the minimum of $H_D$ compared to the hyperspin with no equalization, which instead shows a monotonous decrease of $\langle\delta E\rangle$ with increasing $\delta h$, confirming previous observations~\cite{strinati2022hyperspinmachine}. Furthermore, the increased accuracy in finding the minimum energy of the simulated spin model comes together with a strong decrease of heterogeneity, quantified in panel \textbf{(b)} by $\langle A_{\rm het}\rangle$, corroborating the findings in Sec.~\ref{sec:dynamicsofamplitudeandenergy1}.

\subsection{Finite-size scaling and performance comparison}
\label{sec:dynamicsofamplitudeandenergy3}
We now present our data on the finite-size scaling of the relative energy deviation and heterogeneity degree. We proceed by repeating the same numerical analysis of Sec.~\ref{sec:dynamicsofamplitudeandenergy2} and Fig.~\ref{fig:numericalresults1} for different values of $N$, and by showing $\delta E$ and $A_{\rm het}$ for each $N$ by performing an additional average over the pump amplitude.

The result is shown in Fig.~\ref{fig:numericalresults2}. The double average over graphs and pump amplitude is denoted by $\overline{\langle\delta E\rangle}$ and $\overline{\langle A_{\rm het}\rangle}$ for the relative energy deviation and heterogeneity degree, respectively. We see that $\overline{\langle\delta E\rangle}$ in panel \textbf{(a)} remains fairly constant in $N$ for both the hyperspin machine with no equalization (red data) and the equalized hyperspin machine (blue data). Remarkably, the data of $\overline{\langle\delta E\rangle}$ with equalization are almost two orders of magnitude smaller than those from the hyperspin machine with no equalization for all scanned values of $N$.

Panel \textbf{(b)} shows the finite-size scaling of $\overline{\langle A_{\rm het}\rangle}$. The degree of heterogeneity from the equalized hyperspin machine (blue data) is several orders of magnitude smaller than that from the hyperspin machine with no equalization (red data), at all scanned values of $N$. Moreover, we see that the heterogeneity degree does not considerably vary with $N$ for the hyperspin machine with no equalization, while it varies by three orders of magnitude for the equalized hyperspin machine, in particular increasing with $N$. This result is ascribed to the fact that the relaxation time to converge to the equalized state increases with $N$, as observed in Fig.~\ref{fig:nonlinarmaphistogramdynamics01}, and therefore the limited simulation time $N_{\rm rt}$ is not large enough to allow the actual reach of the equalized state. Nevertheless, by comparing the scaling of $\overline{\langle A_{\rm het}\rangle}$ with that of $\overline{\langle\delta E\rangle}$, we observe that despite $\overline{\langle A_{\rm het}\rangle}$ increases with $N$, the scaling of $\overline{\langle\delta E\rangle}$ is not sensitively affected by the non-optimal convergence to the equalized steady state. This suggests a robustness of the equalized hyperspin machine against non-perfect amplitude equalization.

Before concluding, we stress that the reason why report data up to $N=400$, even if we have successfully tested the equalized hyperspin machine up to $N=10000$ (Fig.~\ref{fig:nonlinarmaphistogramdynamics01}), is because we observed that the equalized hyperspin machine for several graphs at $N>400$ found a slightly lower value of energy compared to $E_{D,{\rm min}}$ determined by python, most probably due to the difficulty of the python minimizer \texttt{scipy.optimize.basinhopping} to find the actual global minimum for increasing $N$. On one hand, since this analysis requires the determination of the minimum energy $E_{D,{\rm min}}$ to benchmark the hyperspin machine, a non-reliable determination $E_{D,{\rm min}}$ prevents us from pushing the finite-size scaling beyond $N=400$. On the other hand, this fact is an indication that the equalized hyperspin machine as a minimizer of $H_D$ offers competitive performance with existing numerical minimizers.


\section{Conclusions}
\label{sec:conclusions}
In this paper, we introduced and demonstrated a method to implement the hyperspin machine with equalized hyperspin amplitudes. The hyperspin machine introduced in Refs.~\cite{strinati2022hyperspinmachine,PhysRevLett.132.017301} is a network of linearly and nonlinearly coupled parametric oscillators, working as a simulator of coupled classical spin vectors, with arbitrary number of components. We first recalled the construction of a single hyperspin, and then reviewed the hyperspin machine, discussing its working principle as a gradient descent system performing unconstrained optimization on the $D$-vector model Hamiltonian. By ``unconstrained'' we mean that the hyperspin amplitudes are free to converge to largely unequal values in the steady state. Similarly to the Ising machine, unconstrained amplitudes cause the system to minimize a cost function that differs from the desired, target $D$-vector model. In the hyperspin machine, this issue translates into a steady-state hyperspin energy that (i) Can be significantly larger than the global minimum of the  $D$-vector model Hamiltonian, and (ii) It's value depends on the system parameters, particularly on the pump amplitude.

To overcome these two issues, we introduced our method to equalize the hyperspin amplitudes during the dynamics, and in particular in the steady state. Our method employs an additional layer of oscillators (named equalizers). The equalizers are connected to the hyperspin machine via an antisymmetric nonlinear coupling, and their role is to ``absorb'' amplitude heterogeneity in the hyperspin machine. As a result, in the steady state, the equalizers amplitudes are strongly unequal, but the hyperspin amplitudes are equalized.

We analytically showed that the existence of the steady state with equal amplitudes is intimately connected to the reach of the minimum of the simulated $D$-vector model Hamiltonian. In order to demonstrate our method on large-scale systems, we focused on a specific class of coupling matrices, describing random graphs with fixed edge density and binary edge weight. We showed that, when the minimum energy of the $D$-vector model and the maximum eigenvalue for a specific graph has a nontrivial scaling with $N$, the coupling weight should be properly scaled in order to ensure the existence of the fixed point with equal hyperspin amplitudes at all scales.

By large-scale parallel numerical simulations based on the formalism of nonlinear maps, we validated the hyperspin machine with amplitude equalization up to $N=10^4$ hyperspins. By comparing the minimum energy value reached by the hyperspin machine with and without amplitude equalization, we clearly showed that amplitude equalization drastically reduces the energy value, confirming that the equalized hyperspin machine minimizes the $D$-vector model Hamiltonian more reliably due to proper cost function mapping.

We then performed a finite-size statistical analysis. For a specific coupling matrix, we compared the energy from the hyperspin machine relative to the minimum energy value (relative energy deviation) found by a python numerical minimizer, and we related its value to the heterogeneity degree. For increasing number of hyperspins $N$, we averaged the relative energy deviation and the heterogeneity degree over $50$ random binary sparse graphs.

Our analysis unveiled that (i) The hyperspin machine with equalized amplitudes consistently finds orders of magnitude lower hyperspin energy compared to the hyperspin machine without amplitude equalization, showing a reduced sensitivity to the pump amplitude, and (ii) The lower minimum hyperspin energy comes together with a reduced amplitude heterogeneity in the system. Remarkably, we find that the ability of the equalized hyperspin machine to find significantly lower energy values compared to its unconstrained counterpart is not drastically affected by the presence of some residual amplitude heterogeneity in the system, whose presence is ascribed mostly to finite simulation times.

We here considered the case of each equalizer connected to all $D$ oscillators forming an hyperspin, but we can easily select only a part of the oscillators $D'<D$ to be connected to the equalization layer. This opens the possibility to combine amplitude equalization with dimensional annealing. As a future perspective, it will be important to verify whether this can further boost the ability of the hyperspin machine with dimensional annealing to work as an Ising minimizer.

In this work we conducted the finite-size scaling analysis up to $N=400$ hyperspins. It is important to remark that this limited scaling was a consequence of the fact that the equalized hyperspin machine for larger values of $N$ was able to find lower value of energy compared to the benchmark energy determined by python numerical minimization. This result further corroborates the validity of the equalized hyperspin machine as a minimizer of the simulated $D$-vector model Hamiltonian.

\begin{acknowledgements}
We acknowledge the CINECA award under the ISCRA initiative, for the availability of high performance computing resources and support. Large-scale numerical simulations have been performed using an extended version of the parallelized C-language code previously developed in Ref.~\cite{PhysRevLett.132.017301} for the hyperspin machine map. 
Funded by the European Union (HORIZON-ERC-2023-ADG HYPERSPIM project grant number 101139828). Views and opinions expressed are however those of the author(s) only and do not necessarily reflect those of the European Union or the European Research Council. Neither the European Union nor the granting authority can be held responsible for them.
\end{acknowledgements}


%

\end{document}